\documentclass[twoside]{aa} 
\usepackage{caption}
\usepackage{graphicx}
\usepackage{setspace}
\usepackage{natbib}
\usepackage{rotating}
\usepackage{threeparttable}
\usepackage{tabto}
\usepackage{float}
\usepackage{subcaption}
\usepackage{amsmath}
\usepackage{nccmath}
\usepackage{orcidlink}
\usepackage{adjustbox}
\usepackage{lscape}
\usepackage{bm}
\usepackage{amsmath,siunitx}
\usepackage{geometry}
\newgeometry{vmargin={15mm}, hmargin={12mm,17mm}} 
\bibpunct{(}{)}{;}{a}{}{,} % to follow the A&A style

\usepackage{txfonts}
\usepackage{hyperref}
\hypersetup{
  colorlinks=true,   %% links colored instead of frames
  urlcolor=blue,     %% external hyperlinks
  linkcolor=blue,     %% internal latex links (eg Fig)
  citecolor=blue
}

\newcommand{\msun}{\mbox{M$_{\odot}$}}
\newcommand{\rsun}{\mbox{R$_{\odot}$}}
\newcommand{\lsun}{\mbox{L$_{\odot}$}}
\newcommand{\zsun}{\mbox{Z$_{\odot}$}}

\providecommand{\sun}{_\text{\sun}}

\DeclareUnicodeCharacter{2212}{\ensuremath{-}}

\makeatletter
\renewcommand*\aa@pageof{, page \thepage{} of \pageref*{LastPage}}
\makeatother

\begin{document} 
    \title{Investigating episodic mass loss in evolved massive stars}
   \subtitle{IV. Comprehensive analysis of dusty red supergiants \\ in NGC 6822, IC 10, and WLM}

   \author{E. Christodoulou \inst{1,2,}\thanks{Corresponding author: \texttt{evachris@noa.gr}} \orcidlink{0000-0003-4332-3646}
   \and S. de Wit \inst{1,2}\orcidlink{0000-0002-9818-4877}
   \and A.Z. Bonanos \inst{1}\orcidlink{0000-0003-2851-1905} 
   \and G. Muñoz-Sanchez \inst{1,2}\orcidlink{0000-0002-9179-6918} 
    \and G. Maravelias  \inst{1,3}\orcidlink{0000-0002-0891-7564} 
   \and A. Ruiz \inst{1}\orcidlink{0000-0002-3352-4383}
  \and \\ K. Antoniadis \inst{1,2}\orcidlink{0000-0002-3454-7958}
   \and D. Garc\'ia-\'Alvarez \inst{4,5}\orcidlink{}    
   \and M.M. Rubio D\'iez \inst{6}\orcidlink{0000-0003-4076-7313}
   }
    
   \institute{
    IAASARS, National Observatory of Athens, 15326 Penteli, Greece
    \and
    National and Kapodistrian University of Athens, Department of Physics, Panepistimiopolis, 15784 Zografos, Greece  
    \and
    Institute of Astrophysics FORTH, 71110 Heraklion, Greece
    \and
    Instituto de Astrofísica de Canarias, Avenida Vía Láctea, 38205 La Laguna, Tenerife, Spain
    \and
    Grantecan S. A., Centro de Astrofísica de La Palma, Cuesta de San José, 38712 Breña Baja, La Palma, Spain
    \and
    Centro de Astrobiolog\'ia, CSIC-INTA, Ctra. de Torrej\'on a Ajalvir km 4, E-28850 Torrej\'on de Ardoz, Madrid, Spain
    }       
   \date{}
   
   \abstract
  {Mass loss shapes the fate of massive stars; however, the physical mechanism causing it remains uncertain. We present a comprehensive analysis of seven red supergiants, for which we searched evidence of episodic mass loss, in three low-metallicity galaxies: NGC~6822, IC~10, and WLM. Initially, the spectral classification of their optical spectra was refined and compared to previous reported classifications, finding four sources that display spectral variability. We derived the physical properties of five of them using the \textsc{marcs} atmospheric models corrected for nonlocal thermal equilibrium effects to measure stellar properties from our new near-infrared spectra, such as the effective temperature, surface gravity, metallicity, and microturbulent velocity. Additional empirical and theoretical methods were employed to calculate effective temperatures, finding consistent results. We constructed optical and infrared light curves, discovering two targets in NGC~6822 with photometric variability between 1 and 2.5 mag in amplitude in $r$ and $\sim$ 0.5 mag in the mid-infrared. Furthermore, we discovered a candidate-dimming event in one of these sources. Periods for three red supergiants were determined using epoch photometry, which were consistent with the empirical estimations from literature period-luminosity relations. Our comprehensive analysis of all the available data for each target provides evidence for episodic mass loss in four red supergiants. 
  }

   \keywords{stars: massive - stars: supergiants - stars: fundamental parameters - stars: atmospheres -stars: mass-loss - stars: late type}

   \titlerunning{Comprehensive analysis of dusty RSGs in NGC~6822, IC~10, and WLM}
   \authorrunning{Christodoulou et al.}

\maketitle

%__________________________________________________________________

\section{Introduction}\label{sec:intro}

Massive stars, which are stars with masses M $\geq$ 8 $\msun$, are powerful engines that drive the evolution of their host galaxies with their feedback \citep[e.g.,][]{McCray1987}. At the end of their lifetime, they explode as a supernova or implode directly to a black hole \citep[][]{Heger2003}. Most of them evolve through the red supergiant (RSG) phase \citep{Levesque2017}, during which they undergo both long timescale mass-losing processes, such as stationary winds \citep[e.g.,][]{Beasor2020, Antoniadis2024}, as well as short timescale events, such as eruptive mass loss \citep{Smith2014, Humphreys2022}. These processes and their potential metallicity dependence remain poorly understood \citep[e.g.,][]{Decin2021}, although they determine how much of the H envelope a star ultimately sheds and therefore its finale \citep[e.g.,][]{Smith2011, Georgy2012}.

To understand the winds of RSGs and their driving mechanism, one needs to model their atmospheres. However, this task poses a significant challenge, as their structure is very complex due to the outer layers being convective \citep{Freytag2002, Ludwig2009}, exhibiting granulation \citep{Chiavassa2009}, as well as strong turbulence \citep[e.g.,][]{Josselin2007, Ohnaka2017} -- phenomena induced by convection \citep{Freytag2012}. To study these effects, 3D magnetohydrodynamical and radiative hydrodynamics simulations are needed \citep[e.g.,][respectively]{Meyer2021, Goldberg2022}. In addition, large-scale convection in the stellar interior and self-excited pulsations result in atmospheric dynamics with strong radiative shock waves \citep{Freytag2024}. These extreme atmospheric conditions make the extraction of their physical parameters, and in particular the effective temperature ($T_{\rm eff}$), quite difficult, while the luminosity (L) can be measured by spectral energy distribution (SED) fitting. The stellar radius (R) is derived from the Stefan-Boltzmann law. 

Various approaches are used to determine the RSG $T_{\rm eff}$; however, some produce discrepant results. The most commonly employed method is to model the titanium oxide (TiO) bands in their optical spectra, as TiO bands are sensitive to $T_{\rm eff}$ and optical spectra are readily available. However, modeling TiO bands yields systematically lower $T_{\rm eff}$ (by $\sim$ 300 K) than other methods \citep[e.g., SED fitting;][]{Davies2013} because these lines form in the outer atmospheric layers of the RSGs, where the optical depth ($\tau$) is less than one, probing lower temperature regions. Furthermore, TiO bands are affected by metallicity and mass loss, which cause them to appear deeper \citep{Davies2021}. The combined effect of $T_{\rm eff}$, metallicity, and mass loss on the strength of the TiO bands cannot be disentangled in the optical.

An alternative method to determine the $T_{\rm eff}$ of RSGs is modeling atomic lines in their $J$-band spectra. The method produces more accurate $T_{\rm eff}$ results than the modeling of the TiO bands because it uses lines that form near the stellar continuum region ($\tau \sim$ 1). This technique was introduced by \citet{Davies2010}, originally aiming to derive metallicities of RSGs by exploiting the fact that their SED peaks in the near-infrared (near-IR), which is applicable even at large distances \citep{Evans2011, Lardo2015}. It comprises of fitting Fe, Si, Ti, and Mg absorption lines present in the observed spectra with synthetic spectra to estimate stellar parameters. \citet{Davies2015} demonstrated the strong $T_{\rm eff}$ sensitivity of the $J$-band modeling method and derived error bars of $\sim$ 50~K, which are comparable with those extracted using the SED fitting technique for the same dataset. This is the most accurate method to calculate RSG $T_{\rm eff}$ \citep[][]{Davies2013}; however, it requires spectroscopic data covering a wide wavelength range. In the absence of $J$-band spectroscopy, empirical or theoretical relations based on $J-K_{s}$ colors can be exploited to derive $T_{\rm eff}$ estimations \citep[e.g.,][]{Neugent2012,deWit2024}. Scaling relations between the TiO temperatures and the $J$-band and $i$-band temperatures, developed by \citet{deWit2024}, can be used to derive more realistic $T_{\rm eff}$ based on the available TiO temperatures. 

Spectral changes (i.e., changes in the TiO bands), associated with $T_{\rm eff}$ variations, can be used to trace RSG mass loss \citep{Davies2021}. RSGs can display spectroscopic variability because of instability phases during which they become more luminous and cooler \citep[e.g.,][]{Levesque2007, Dorda2021}. The mean change reported is two spectral subtypes \citep{Dorda2016}. More extreme cases are the Levesque--Massey variables, which transition from late to early M-type classifications \citep[e.g.,][]{Levesque2009} or between late M-type to early K-type spectra on a timescale of a few months \citep[e.g.,][]{Massey2007}. This variation in the TiO bands can be caused by either episodic mass loss \citep[e.g.,][]{Montarges2021, Munoz-Sanchez2024} or hysteresis loops \citep{Kravchenko2019, Kravchenko2021}. 

Signs of mass loss can also be imprinted onto the RSG light curves. RSGs are known to be semi-regular variables, showing variability of up to a few magnitudes in the optical \citep[][]{Kiss2006}. Luminous RSGs even exhibit significant variability in the mid-infrared \citep[mid-IR; e.g.,][]{Yang2018}. Therefore, epoch photometry from numerous surveys that continuously monitor the sky can be employed to construct detailed light curves useful for detecting episodic mass-loss events that manifest as dimming events \citep[e.g., Betelgeuse, RW~Cep and $\rm{[W60]}$~B90;][]{Guinan2019, Montarges2021, Anugu2023, Kasikov2025, Munoz-Sanchez2024}. These episodic mass ejections are linked to asymmetric gaseous outflows associated with surface activity \citep{Humphreys2022}. 

In this paper, we present a comprehensive study of some of the brightest RSGs identified by the ASSESS project \citep{Bonanos2024} in NGC~6822 and IC~10 \citep{deWit2025}, as well as a RSG reported in the literature in WLM, in search of sources that have undergone episodic mass loss. ASSESS aimed to determine the role of episodic mass loss and surveyed over 1000 evolved massive star candidates in 25 nearby galaxies, yielding eight new RSGs in the Magellanic Clouds \citep{deWit2023}, 129 RSGs in ten other southern galaxies \citep{Bonanos2024}, and 28 RSGs in three northern galaxies \citep{deWit2025}. The paper is structured as follows: in Sect.~\ref{secObs} we describe the observations and the data reduction process. In Sect.~\ref{secDataAn} we discuss our methodology. In Sect.~\ref{secRes} we present our results. In Sect.~\ref{secDis} we discuss the implications of our results and in Sect.~\ref{sec:summary} we summarize our findings and present our conclusions.

\section{Observations} \label{secObs}

\subsection{Target selection}\label{sec:target_sel}

\hfill
\begin{figure*}[h]
   \begin{subfigure}{1.0\columnwidth}
    \centering
    \includegraphics[width=1\linewidth]{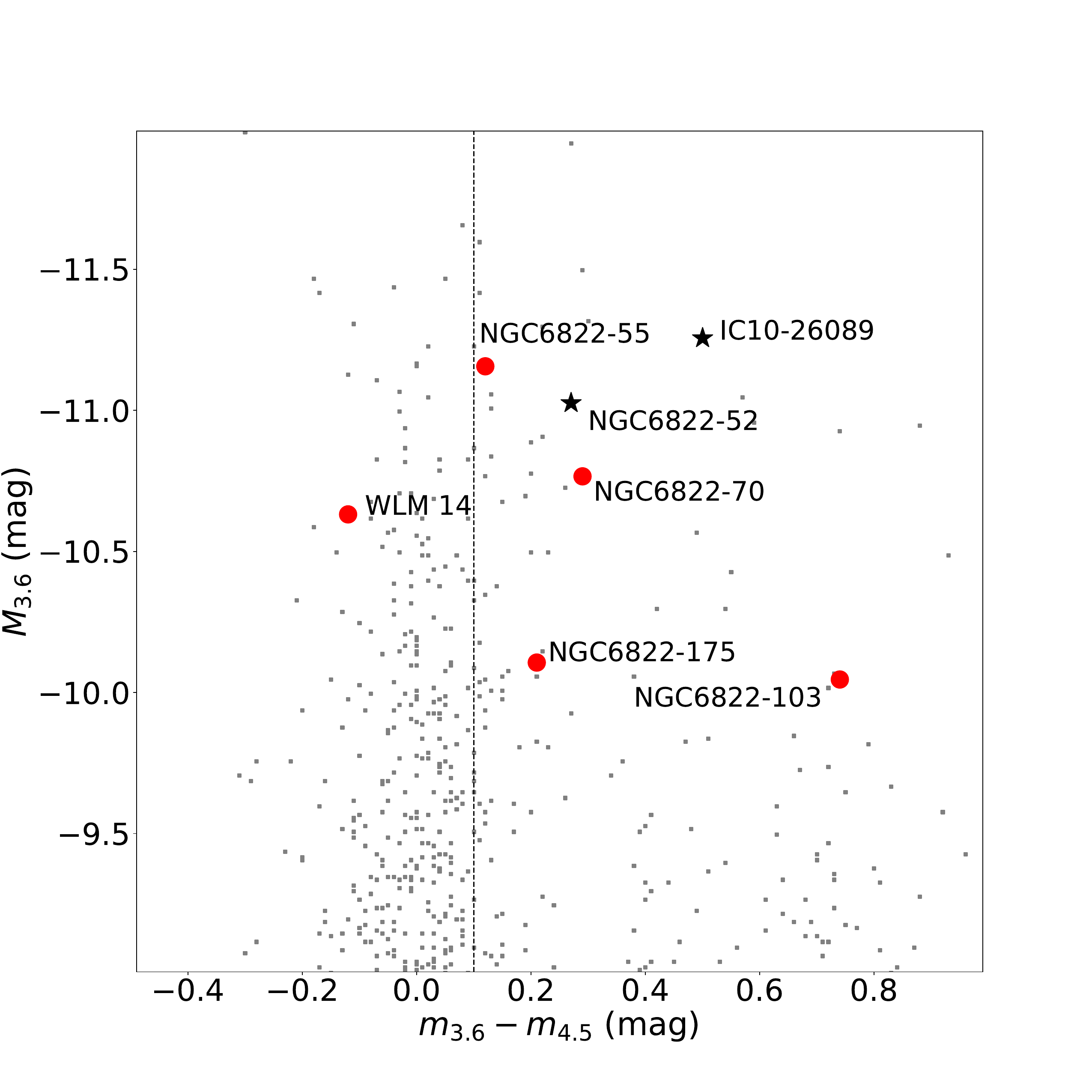}
   \end{subfigure}
\hfill % <--- 
    \begin{subfigure}{1.0\columnwidth}
        \centering
        \includegraphics[width=1\linewidth]{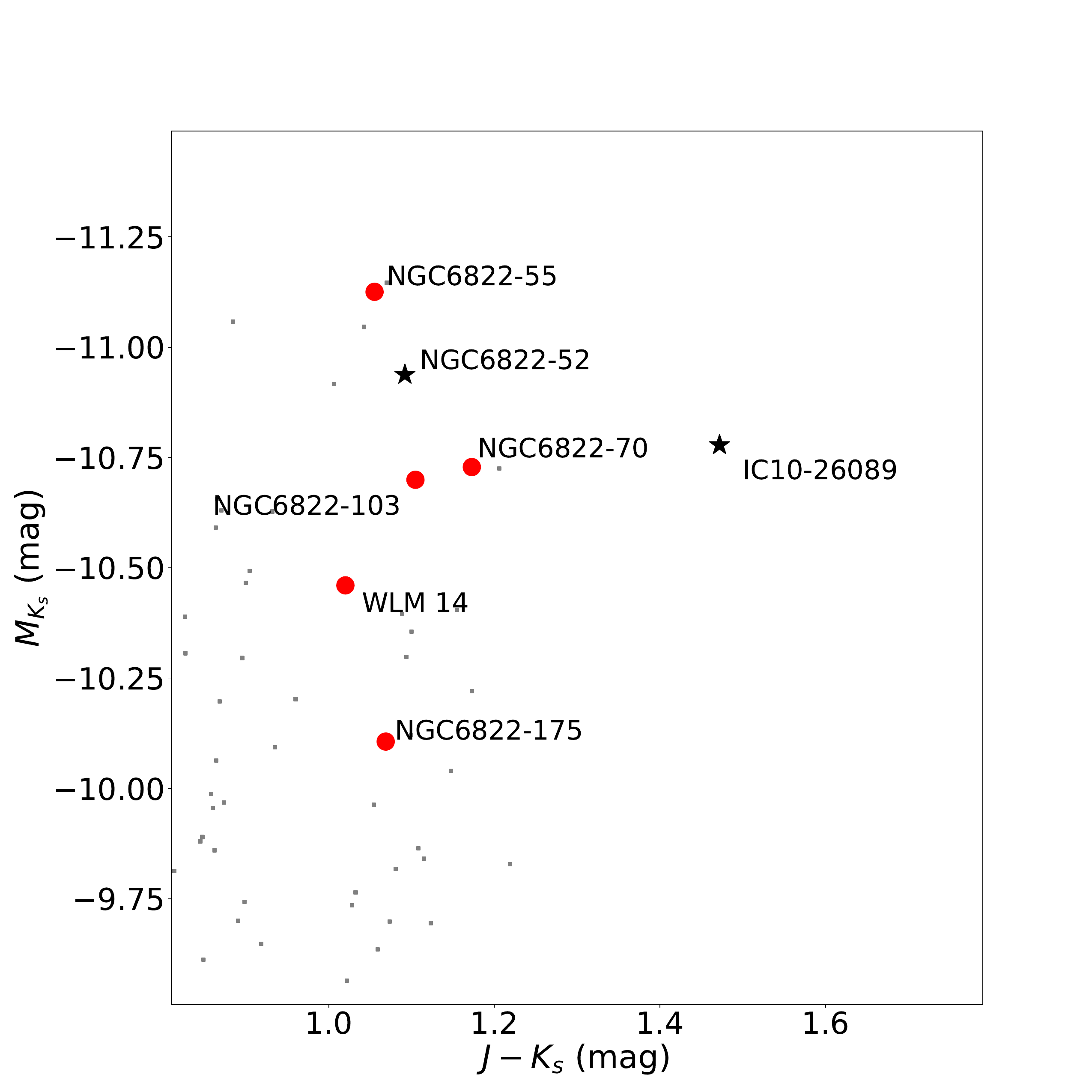}
    \end{subfigure}

\caption{Mid-IR (\textit{left}) and near-IR (\textit{right}) CMDs for the seven targets in NGC~6822, IC~10 and WLM. Red circles indicate targets whose spectra we were able to model. Background sources are stars from NGC~6822. The vertical line in the mid-IR CMD indicates the color criterion used for selecting stars with IR-excess.}
\label{fig:cmd}
\end{figure*}

\begin{figure}[h]
    \centering
\includegraphics[width=1\columnwidth]{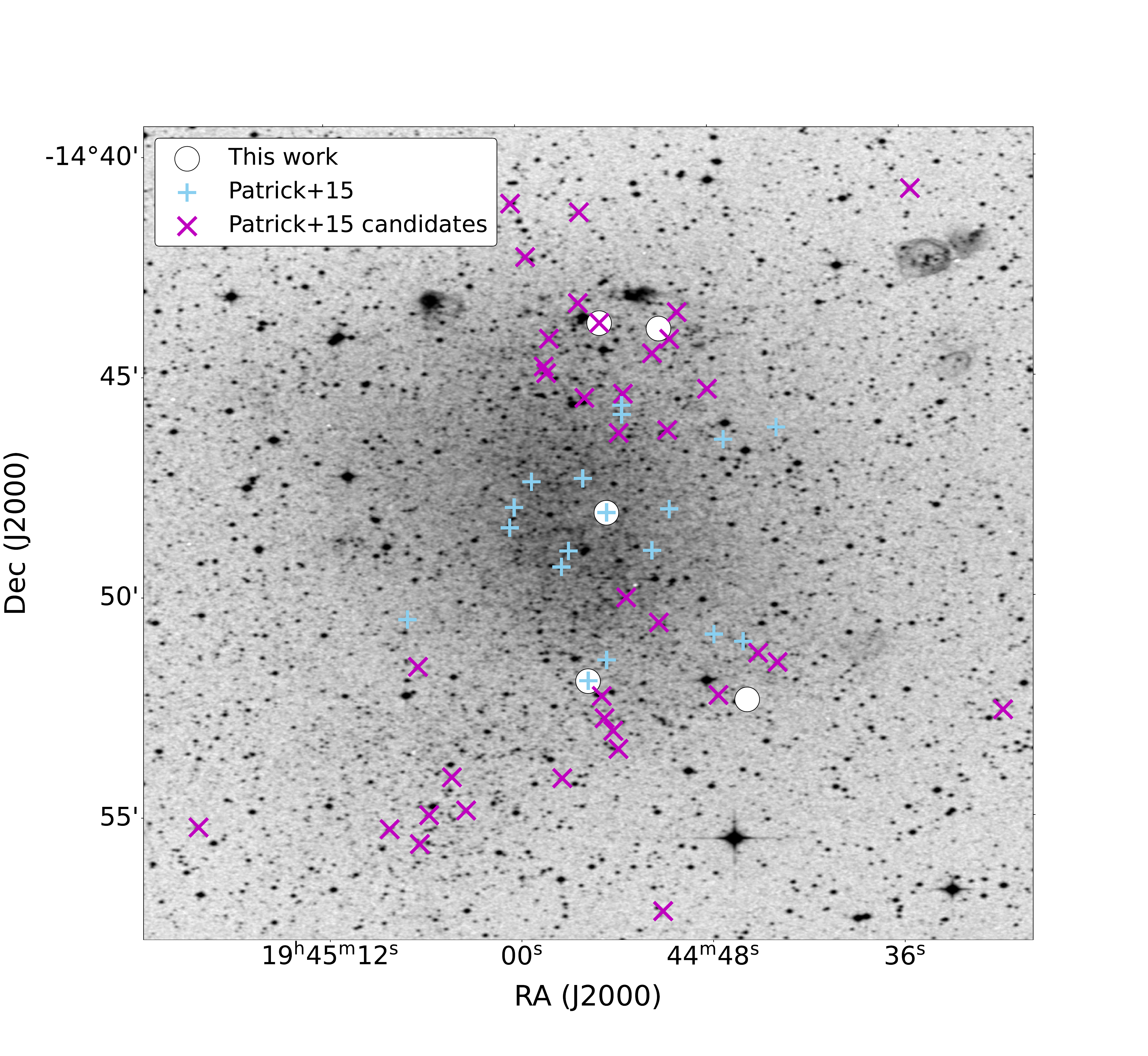}
    \caption{Spatial distribution of our targets in NGC~6822 and the targets of \citet{Patrick2015}, two of which we have also observed. We included RSG candidates from the same work, one of which we have observed (see text for details). The background image is from the DSS2 survey.}
    \label{ngc6822}    
\end{figure}

We selected six of the reddest and most dusty spectroscopically confirmed RSGs \citep[5 in NGC~6822 and one in IC~10 from the study of][]{deWit2025} with available optical spectra from the Optical System for Imaging and low-Intermediate-Resolution Integrated Spectroscopy (OSIRIS) on Gran Telescopio Canarias (GTC). These rare sources \citep[][]{deWit2024}, were chosen using their $J$-band brightness (> 16.0 mag). 
We included two targets, namely NGC6822-55 and NGC6822-175, that have previously been observed in the near-IR \citep{Patrick2015} to perform an independent modeling of their $J$-band spectra to retain a homogeneous approach for all of our targets and investigate potential differences due to spectral variability. We note that NGC6822-70 is included in a catalog of RSG candidates in NGC~6822 compiled by \citet{Patrick2015} (see their section 2.1 for details), which they shared with us (Lee Patrick, priv. comm.). We included NGC6822-52, which is classified as a RSG candidate in \citet{deWit2025}, as its RSG status is confirmed by other studies \citep{Humphreys1980, Massey1998}. Finally, we added a bright, spectroscopically confirmed (in the optical) RSG target, WLM~14 \citep{Britavskiy2019}, to extend our RSG sample to lower metallicities. 
Table~\ref{tab:gal_prop} presents the properties of the host galaxies, such as coordinates, morphological type, distance, mean metallicity, and radial velocity. In Table~\ref{tab:tar_prop}, we present the coordinates, selected photometry, and spectral type both from the literature (determined using optical spectroscopy) and this work for all our targets. 

We constructed color-magnitude diagrams (CMDs; Fig.~\ref{fig:cmd}), from which it is evident that almost all of our targets show IR excess (i.e., those with $m_{3.6}$ – $m_{4.5}$ > 0.1 mag), implying a mass-loss history. Notably, in the near-IR CMD, sources located in different galaxies have different colors. This is probably caused by foreground extinction, which is not significant in the mid-IR. In Fig.~\ref{ngc6822} we present the spatial distribution of our targets in NGC~6822, along with the RSGs observed by \citet{Patrick2015} in the $J$-band and the RSG candidates from the same work. We cross-matched this catalog with the catalog of RSG candidates from \cite{Antoniadis2025} in this galaxy, which was cleaned from foreground contamination using \textit{Gaia} astrometry. We included only candidates present in both catalogs to remove potential foreground sources.

\begin{table*}[h]
    \centering
    \caption{Properties of the host galaxies.}
    \begin{tabular}{l c r c c c r}
    \hline
    \hline
        Galaxy & R.A.  & Dec. & Type \tablefootmark{a}  & Distance$^1$ & Z \tablefootmark{b} & $v_{\rm{rad}}$$^{5,}$ \tablefootmark{c}  \\
        & \small{(J2000)} & \small{(J2000)} &  & \small{(Mpc)} & \small{($\zsun$)} & \small{(km~s$^{-1}$)} \\
        \hline
         IC~10& 00 20 17.3 & +59 18 13.6 & dIrr IV/BCD & 0.74 $\pm$ 0.04 & 0.45$^{2}$ & $-$348 $\pm$ 1 \\
         NGC~6822& 19 44 57.7 & 14 48 12.4 & IB(s)m & 0.48 $\pm$ 0.03 & 0.32$^{3}$ &$-$57 $\pm$ 2 \\
         WLM & 00 01 58.2 & $-$15 27 39.3 & IB(s)m  & 0.96 $\pm$ 0.03 & 0.14$^{4}$ & $-$130 $\pm$1 \\
         \hline
    \end{tabular}
    \tablebib{
    (1) \cite{Tully2013} ; 
    (2) \cite{Tehrani2017}; 
    (3) \cite{Dopita2019} ;
    (4) \cite{Urbaneja2008} ;
    (5) \cite{McConnachie2012} 
    }
    \tablefoot{
\tablefoottext{a}{Retrieved from NASA/IPAC Extragalactic Database (NED).}
\tablefoottext{b}{Oxygen abundances were converted to metallicity using \citet{Asplund2009}. The numbers reported here represent mean values per galaxy.} 
\tablefoottext{c}{The errors of the $v_{\rm{rad}}$ values correspond to the statistical error.}
}
    \label{tab:gal_prop}
\end{table*}

\begin{table*}[h]
\small
\caption{Properties of the targets.} \label{tab:tar_prop}
\begin{tabular}{l r r c c c c c c c c l}
\hline
\hline
ID & R.A. & Dec. & $m_{3.6}$\tablefootmark{b} & $m_{4.5}$\tablefootmark{b} & $J$\tablefootmark{b} & $K_{s}$\tablefootmark{b} & $G$\tablefootmark{b} & $G_{\rm{BP}}$\tablefootmark{b} & $G_{\rm{RP}}$\tablefootmark{b} & Sp. Type\tablefootmark{c} & Sp. Type \\
& \small{(\degr)} & \small{(\degr)} &  \small{(mag)} & \small{(mag)} & \small{(mag)} & \small{(mag)} & \small{(mag)} & \small{(mag)} & \small{(mag)} & \small{This work} & \small{Literature}   \\
\hline
IC10-26089 & 5.02108 & 59.30108 & 13.09 & 12.59 & 15.61 & 14.13 & 19.21 & 20.80 & 17.86 & K4 I & -   \\
NGC6822-52 & 296.21295 & $-$14.73218 & 12.38 & 12.11 & 13.56 &12.47 & 17.07 & 18.72 & 15.82 & M4 I & M2.5 I $^{1}$, M1 I $^{2}$ \\
NGC6822-55\tablefootmark{a} & 296.23214 & $-$14.86554 & 12.25 & 12.13 & 13.34 & 12.28 & 16.20 & 17.48 & 15.09 & K7 I & M0 I  $^{3}$, M1-2 I $^{2}$  \\
NGC6822-70 & 296.22841 & $-$14.73003 & 12.64 & 12.35 & 13.85 & 12.68 & 17.00 & 18.45 & 15.80 & M4 I & cM/RSG $^{1,2}$ \\ 
NGC6822-103 & 296.19068 & $-$14.87268 & 13.36 & 12.62 & 13.81 & 12.71 & 16.66 & 17.90 & 15.57  & M0 I & M1 I $^{3}$  \\
NGC6822-175\tablefootmark{a} & 296.22696 & $-$14.80182 & 13.30 & 13.09 & 14.37 & 13.30 & 17.22 & 18.52 & 16.10 & M4 I & M1 I $^{3}$\\ 
WLM~14 & 0.51268 & $-$15.50950 & 14.28 & 14.40 & 15.43 & 14.45 & 17.74 & 18.73 & 16.77 & - & K4-5 I $^{4,5}$, K5 I $^{3}$ \\
\hline
\end{tabular}
\tablebib{
    (1) \citet{Massey1998} ; 
    (2) \citet{Humphreys1980} ;
    (3) \citet{Levesque2012} ;
    (4) \citet{Britavskiy2019}; 
    (5) \citet{Britavskiy2015}}
\tablefoot{
\tablefoottext{a}{Studied in the $J$-band by \cite{Patrick2015}.}
\tablefoottext{b}{\textit{Spitzer} $m_{3.6}$ and $m_{4.5}$ data have a mean error of 0.04~mag. $J$, $K_{s}$ photometry for sources in NGC~6822 and WLM is from the Vista Hemisphere Survey DR5 \citep{McMahon2013} and 2MASS \citep{Skrutskie2006} for our IC~10 sources. Mean errors in both bands are 0.05 mag. \textit{Gaia} DR3 $G$, $G_{\rm BP}$, $G_{\rm RP}$ photometry data have a mean error of 0.01, 0.07 and 0.03 mag, respectively.}\\
\tablefoottext{c}{See Sect.~\ref{sec:opt_spec}.} \\}
\end{table*}

\subsection{EMIR spectroscopy}\label{sec:emir}

The observations were carried out using Espectrógrafo Multiobjeto Infra-Rojo \citep[EMIR;][]{Garzon2022} at the GTC under program 92-GTC77/22B (PI: D. Garc\'ia-Alvarez) in the $J$-band ($\lambda\lambda$ 11710--13270 \AA). We obtained spectra with an ABBA nodding pattern, where A is the on-object pointing and B is the on-sky pointing. We used the long-slit mode and a slit width of $\ang{;;1.2}$, achieving a resolution of $\sim$3200. The observation log is provided in Table \ref{tab:log}. We report the ID, the coordinates, the UT date of the observation, the total exposure time ($t_{\rm{exp}}$), the airmass, seeing conditions, the moon illumination, and the achieved signal-to-noise ratio (S/N) per object. We observed telluric standard stars, one for each galaxy. The observations took place from September to December 2022, before the upgrade of the instrument in late 2023, meaning that they have been affected by the inefficiency of the old detector \citep{Garzon2024}. To ensure data acquisition, we relaxed the seeing constraints, which notably reduced the S/N compared to our original expectations. 

\subsection{Data reduction}\label{sec:data_reduction}

We used the EMIR dedicated pipeline (PyEMIR) version 0.17.0 for the image rectification, dark and flat-field correction, sky background subtraction, combination of the nodding frames using sigma clipping, and the wavelength calibration. The spectra were extracted with the IRAF\footnote{IRAF is distributed by the National Optical Astronomy Observatory, operated by the Association of Universities for Research in Astronomy (AURA) under agreement with the National Science Foundation.} task \textsc{apall}; cosmic rays were removed with \textsc{lacosmic} \citep{VanDokkum2001}. We used \textsc{molecfit} \citep[version 4.3.1;][]{Smette2015} to correct for telluric features. Its implementation is discussed in detail in Sect.~\ref{sec:tell_corr}. We note that PyEMIR gives vacuum wavelengths and does not perform the heliocentric correction. We calculated and performed the heliocentric correction at a later stage (see Sect.~\ref{sec:fit_process}).

We had two partially observed targets, NGC6822-52 and NGC6822-55. The observations were completed, but the objects were detected only in nod A. For these, we extracted the spectrum solely from adding those frames and removed the background manually.
Furthermore, in the case of IC10-26089, we could not use all of the observed nodes as the last block was incomplete. We discarded it and used only the complete blocks. Finally, spectra with S/N lower than 10 were excluded, resulting in five (out of seven) spectra for further analysis. 

\subsubsection{Telluric correction}\label{sec:tell_corr}

We implemented \textsc{molecfit} on EMIR observations (EsoRex version 3.13.7) to correct the telluric features. \textsc{molecfit} is used to determine a synthetic spectrum of the Earth's atmosphere, based on local weather conditions and standard atmospheric profiles. It utilizes a radiative transfer code and a molecular line database. It also models the instrumental line spread function. After the instrument features are taken into account, the synthetic transmission spectrum is used to correct the observed spectrum for telluric features.
For the targets in NGC~6822, we utilized the science spectrum of each source directly to calculate the parameters needed to extract the atmospheric model, since the telluric standard star observation for this galaxy was incorrect (nods were obtained in the wrong order). In contrast, for the WLM target, we used the spectrum of the appropriate telluric standard star to obtain the atmospheric model, since it was observed correctly. We then manually divided the star's spectrum with the model to get the telluric corrected spectrum. The parameters relevant to the fit and specific to EMIR are provided in Table~\ref{tab:mol_par}.

\section{Analysis} \label{secDataAn}

\subsection{Optical spectroscopy}\label{sec:opt_spec}
\begin{figure}
\centering
\includegraphics[width=1.0\columnwidth]{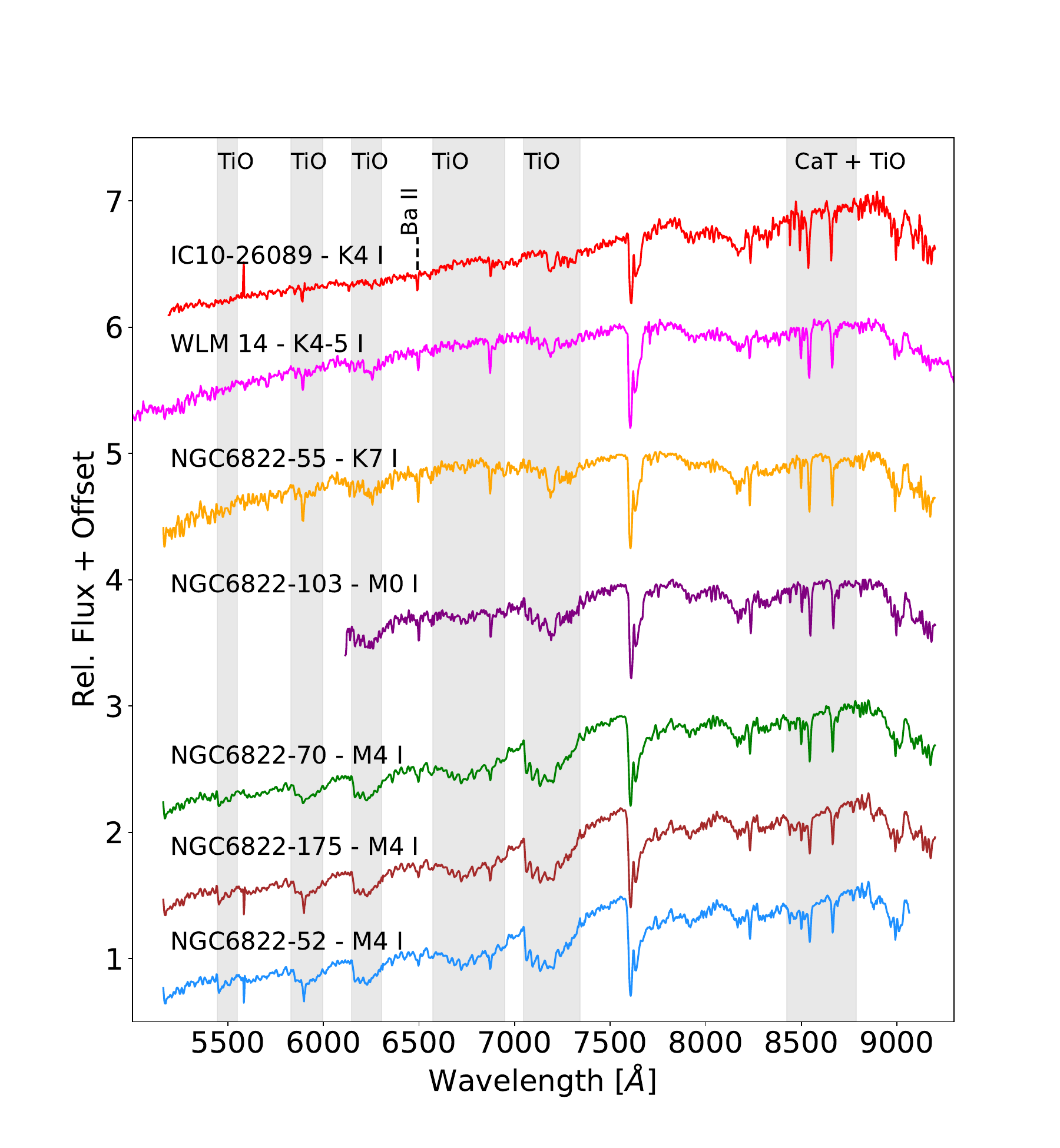}
\caption{Revized spectral types of OSIRIS targets from \citet{deWit2025}, ordered in sequence of spectral type. Important spectral features are indicated. The spectrum and classification for WLM~14 are from \citet{Britavskiy2019}.}
\label{optical_spectra} 
%}
\end{figure}

We reexamined and improved the classification of the optical spectra of the sources we have in common with \citet{deWit2025}, determining them to one subtype. We used the spectral libraries of \citet{Levesque2005}, \citet{Levesque2006}, and \citet{Rayner2009}, since they include classified RSG spectra in a similar wavelength range to \citet{deWit2025}. In Fig.~\ref{optical_spectra}, we present the spectra in order of the strength of the TiO bands (earliest to latest spectral type), from the weakest bands (IC10-26089) on the top to the strongest bands (NGC6822-52) at the bottom. We find a spectral type range from K4 to M4. Crucial spectral features for the classification are highlighted. For completeness, we include the optical spectrum and spectral classification obtained by \citet{Britavskiy2019} for the WLM target. 

We assigned a K4 spectral type to IC10-26089 because the TiO bands are either weak or completely absent. We classified NGC6822-55 as K7 because the TiO bands are deeper, yet still not well-formed. We assigned an M0 spectral type to NGC6822-103 because the TiO bands have become apparent, but the one at 7054~\AA~is not strong. Finally, we classified NGC6822-70, NGC6822-175, and NGC6822-52 as M4 (the latest spectral type) because the TiO band at 6658~\AA~is distinct; however, it is not strong. Therefore, three (out of six) of our dusty RSGs are late-type (therefore evolved) RSGs, reproducing the properties of the dusty RSG population reported by \citet{deWit2024}.

\subsection{The grid} \label{sec:grid}

We constructed our model grid using 1D \textsc{marcs} models \citep{Gustafsson2008} for RSGs, corrected for nonlocal thermal equilibrium effects \citep{Bergemann2012, Bergemann2013, Bergemann2015}, which are publicly available from \citet{NLTE_MPIA}. The models were convolved to match the resolution of our observations using \textsc{PyAstronomy.instrBroadGaussFast}. Next, the wavelength grid of the models was resampled to match the grid of our spectra, and the vacuum wavelengths of our spectra were converted to air wavelengths. We fit the models to the observations to estimate $T_{\rm eff}$, surface gravity ($\log \it{g}$), metallicity ([Fe/H]) and microturbulent velocity ($\xi$). The grid parameters span the following ranges: $3300\leq T_{\rm eff} \leq 4500$~K with a step of 100~K, $-0.5 \leq \log \it{g} \leq$ 0.5 with a step of 0.1, $-1.0 \leq $ [Fe/H] $ \leq 0.0$ with a step of 0.1 and $2.0 \leq \xi \leq 6.0$ with a step of 0.5, forming a grid of 14,157 models. The metallicity range selected was based on the metallicities of RSGs located in NGC~6822 measured by \citet[][see their Table 4]{Patrick2015} 
and on the metallicities of blue supergiants in WLM determined by \citet[][see their Table 4]{Urbaneja2008}. We fit for the radial velocity (RV) prior to the application of the heliocentric correction.

\subsection{Spectral fitting}\label{sec:fit_process}
We fit $\sim$8 selected spectral windows of $\sim$15 \AA\ on average, containing crucial spectroscopic diagnostics. Notably, adjacent diagnostics are included in the same spectral window. Consequently, regions without diagnostics did not influence the resulting best-fit $\chi^2$. There were cases where we could not use all the windows because of the presence of artifacts or noise contaminating the line profiles. Due to the low S/N, there was significant uncertainty in the flux, and the estimated $\chi^2$ was similar for all the models, making it impossible to determine the best-fit model.

To address this issue, we devised a method utilizing \textsc{Ultranest} \citep{Buchner2021}, which is a Bayesian workflow package that uses nested sampling \citep{Skilling2004} to obtain the best model (the one with the largest marginal likelihood) for each star. We assumed a uniform prior for each parameter. \textsc{Ultranest} then computed the posterior distributions for each free parameter and their pair-wise correlations. We renormalized continuum regions of $\sim$10~\AA\ width at the left and the right side of the spectral windows to achieve the best normalization possible, facilitating the fitting process. The exact size in \AA\ of each region depends on the proximity of the spectral windows to each other. We adjusted their span so that their sum around each window would be $\sim$~20~$\AA$. When there was insufficient margin on one side due to the proximity of another spectral feature, we included only the region on the other side. During this process, we removed outliers with values above and below 10\% of the median value of the flux in each region. The process was applied to the five spectra with S/N > 10. After the fitting process, we performed the heliocentric correction and report the final value of the radial velocity ($v_{\rm{rad}}$) in Table~\ref{tab:fit_par}.

\begin{table*}[t]
\small
\centering
\caption{Physical parameters of our targets determined from spectral fitting.}
\label{tab:fit_par}
\begin{tabular}{l l l l l l}
\hline
\hline
        Object & $T_{\rm{eff}}$ (K) & $\log$ \it{g} & [Fe/H] & $\xi$ (km s$^{-1}$) & $v_{\rm{rad}}$ (km s~$^{-1}$)  \\
        \hline
         NGC6822-55 & 4286$^{+136}_{-145}$  & 0.1$^{+0.3}_{-0.4}$ & $-$0.50$^{+0.11}_{-0.11}$ & 5.8$^{+0.1}_{-0.3}$ & $-$43$^{+3}_{-4}$   \\
         NGC6822-70 & 3688$^{+153}_{-124}$  &  0.3$^{+0.1}_{-0.2}$ & $-$0.83$^{+0.19}_{-0.11}$ & 5.2$^{+0.5}_{-0.6}$& $-$39$^{+3}_{-3}$   \\
         NGC6822-103 & 3845$^{+115}_{-108}$  & 0.3$^{+0.2}_{-0.3}$  & $-$0.54$^{+0.14}_{-0.14}$ & 5.2$^{+0.5}_{-0.5}$ & $-$81$^{+3}_{-3}$  \\
         NGC6822-175 & 4383$^{+80}_{-108}$ & 0.3$^{+0.1}_{-0.3}$ & $-$0.81$^{+0.11}_{-0.10}$ & 5.6$^{+0.3}_{-0.5}$ & $-$43$^{+4}_{-4}$   \\
         WLM~14 & 4349$^{+104}_{-233}$ & 0.1$^{+0.3}_{-0.4}$ &
         $-$0.88$^{+0.15}_{-0.09}$ &
         4.6$^{+0.8}_{-1.0}$ & $-$48$^{+6}_{-7}$   \\
         \hline
\hspace*{-4.0cm}
\end{tabular}
\tablefoot{Reported uncertainties are 68.3\% confidence level errors.}       
\end{table*}

\subsection{Other methods to determine effective temperature}\label{sec:teff_other}

We employed both theoretical and empirical $T_{\rm eff}$$(J-K_{s})$ relations to determine $T_{\rm eff}$ for our targets at the epoch of the near-IR photometry, to compare with the results from spectroscopy and trace any $T_{\rm eff}$ variations. These relations require the $J-K_{s}$ color to be dereddened. Therefore, we applied the $E(B-V)$ values determined by \citet{deWit2025} from modeling the optical spectra of our targets with \textsc{marcs} models. We followed the same color correction process outlined by \citet[][see their Sect. 4.2 for details]{deWit2024}. However, the $E(B-V)$ values derived by this method are underestimated \citep[see][for details]{Davies2013}, therefore the $(J-K_{s})_{0}$ colors  are likely upper limits.

We used the empirical $T_{\rm eff}(J-K_{s})_{0}$ relation for the $J$-band, derived by \citet[][see their Table 5]{deWit2024}, to compute $T^{J-K_{s}}_{\rm eff}$. Additionally, we utilized the theoretical $T_{\rm eff}(J-K_{s})_{0}$ relation from the synthetic photometry of the \textsc{marcs} models for the metallicity range $−$1.0 to $+$0.25 dex, derived in the same work, to calculate $T^{\rm MARCS}_{\rm eff}$. Finally, we applied the scaling relation for the $J$-band presented in the same work to derive the predicted $J$-band temperatures from the TiO band temperatures obtained by \citet{deWit2025}. 
The $T_{\rm eff}$ values calculated using all methods described above are presented in Table~\ref{tab:teff}. 

\subsection{Light
curves}\label{sec:lc}

We constructed light curves for all targets in Table~\ref{tab:tar_prop}, by collecting $r,g$-band photometry from the Zwicky Transient Facility \citep[ZTF;][]{Bellm2019, Graham2019}, $o$ and $c$ forced photometry from the Asteroid Terrestrial-impact Last Alert System \citep[ATLAS;][]{Tonry2018, Heinze2018, Shingles2021}, $G$, $BP$, $RP$-band photometry from \textit{Gaia} \citep{GaiaDR12016, GaiaDR32023}, $griyz$-band photometry from the Panoramic Survey Telescope and Rapid Response System \citep[Pan-STARRS1;][]{Wainscott2016,Chambers2016}, near-IR $JHK_{s}$-band photometry \citep{Nagayama2003} and $W1$ and $W2$-band photometry from \textit{NEOWISE} \citep{Mainzer2011}. For the objects located in NGC~6822 and IC~10, the data were taken from \citet{deWit2025}, except for the \textit{NEOWISE} data, which were obtained using the Python module developed by \citet{Hwang2020}.\footnote{The Python module is available on GitHub: \url{https://github.com/HC-Hwang/wise_light_curves}} Near-IR light curves of our NGC 6822 targets are from \citet{Whitelock2013}. For WLM~14, we collected the data from the same surveys following the process described by \citet{deWit2025}, except for the \textit{NEOWISE} data, which were obtained as described above. 

\begin{figure*}[h]
    \centering    \includegraphics[width=2.0\columnwidth]{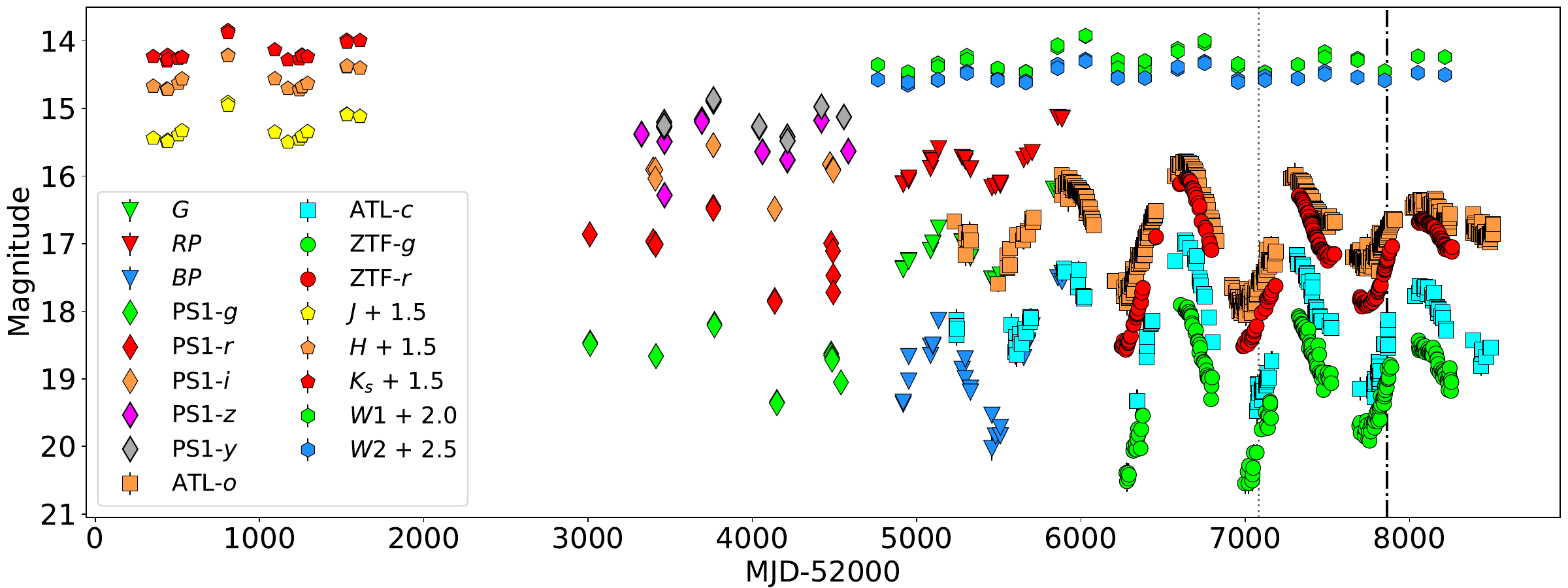}
    \caption{Light curve of NGC6822-52. Filters and surveys are specified in the legend. Vertical lines indicate the epochs spectroscopy was obtained for this source. The dash-dotted line represents the epoch of our EMIR observations, while the dotted line denotes the epoch of the OSIRIS data \citep{deWit2025}.}
    \label{lc_ngc6822-52}    
\end{figure*}

We binned our ZTF, ATLAS, and NEOWISE data from the same night by calculating the median measurement per night, following \citet{deWit2025}. The light curve of NGC6822-52 shows remarkable variability of 2.5 mag and is presented in Figure~\ref{lc_ngc6822-52}. The vertical lines indicate the epochs when spectroscopy was taken. The light curves for the rest of the targets are shown in Appendix~\ref{app:lc}. Next, we performed an analysis of the periodicity of our targets. We used $r,g$-band data from ZTF to construct Lomb--Scargle (L-S) periodograms \citep{Lomb1976, Scargle1982} and present the results of this process in Sect.~\ref{sec:phot_res}.

\section{Results} \label{secRes}

\subsection{Spectral fitting results} \label{specres}

\begin{figure}
    \centering
    \includegraphics[width=1.0\columnwidth]{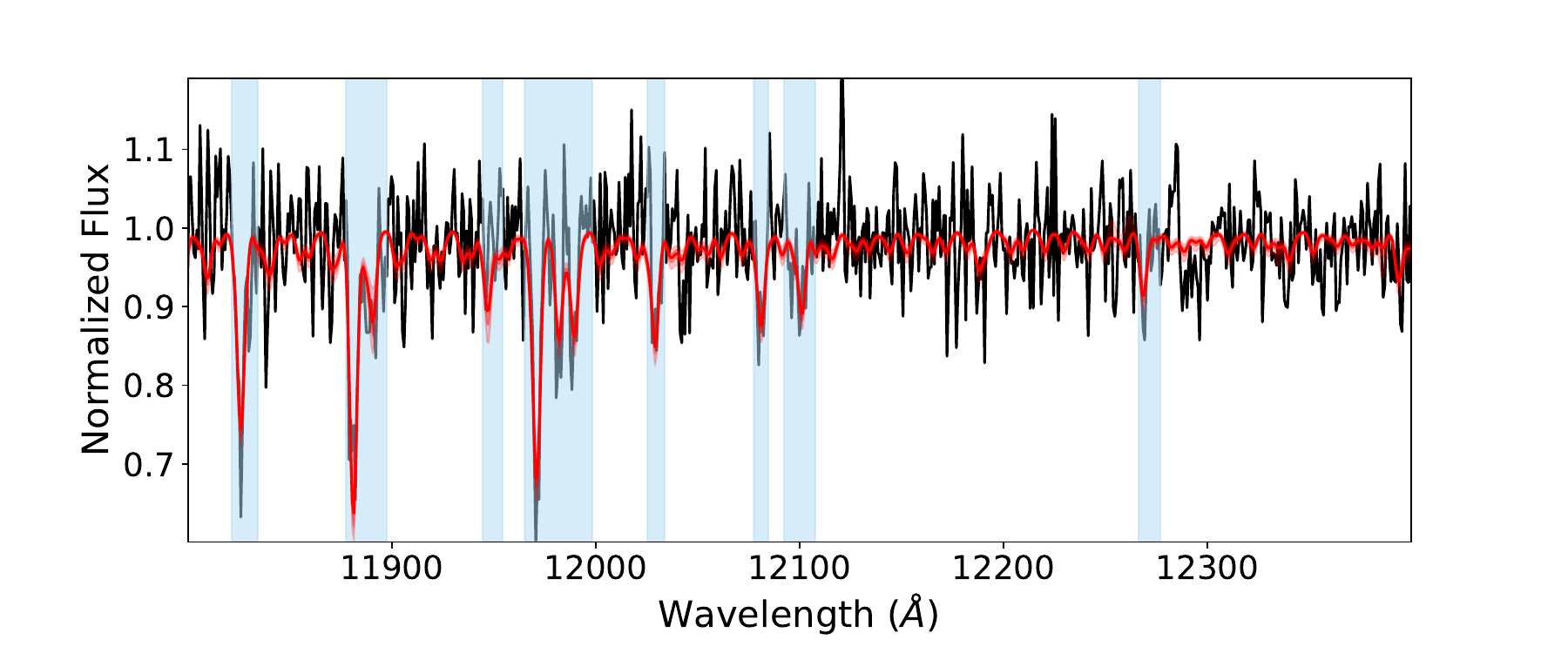}
    \caption{Observed spectrum of NGC6822-103 (in black) and best-fitting model (in red). The shaded blue regions highlight the spectral windows considered for the fitting process. The lines used for the fitting process from left to right by species are Fe~\textsc{i} $\lambda$$\lambda$ 11882.847, 11973.050; Ti~\textsc{i} $\lambda$$\lambda$ 11892.878, 11949.542; Si~\textsc{i} $\lambda$$\lambda$ 11984.20, 11991.57, 12031.50, 12103.54, 12270.50; Mg~\textsc{i} $\lambda$$\lambda$ 11828.185, 12083.346.}
    \label{fit_ngc6822-103}    
\end{figure}

\begin{figure}
    \centering
    \includegraphics[width=1.0\columnwidth]{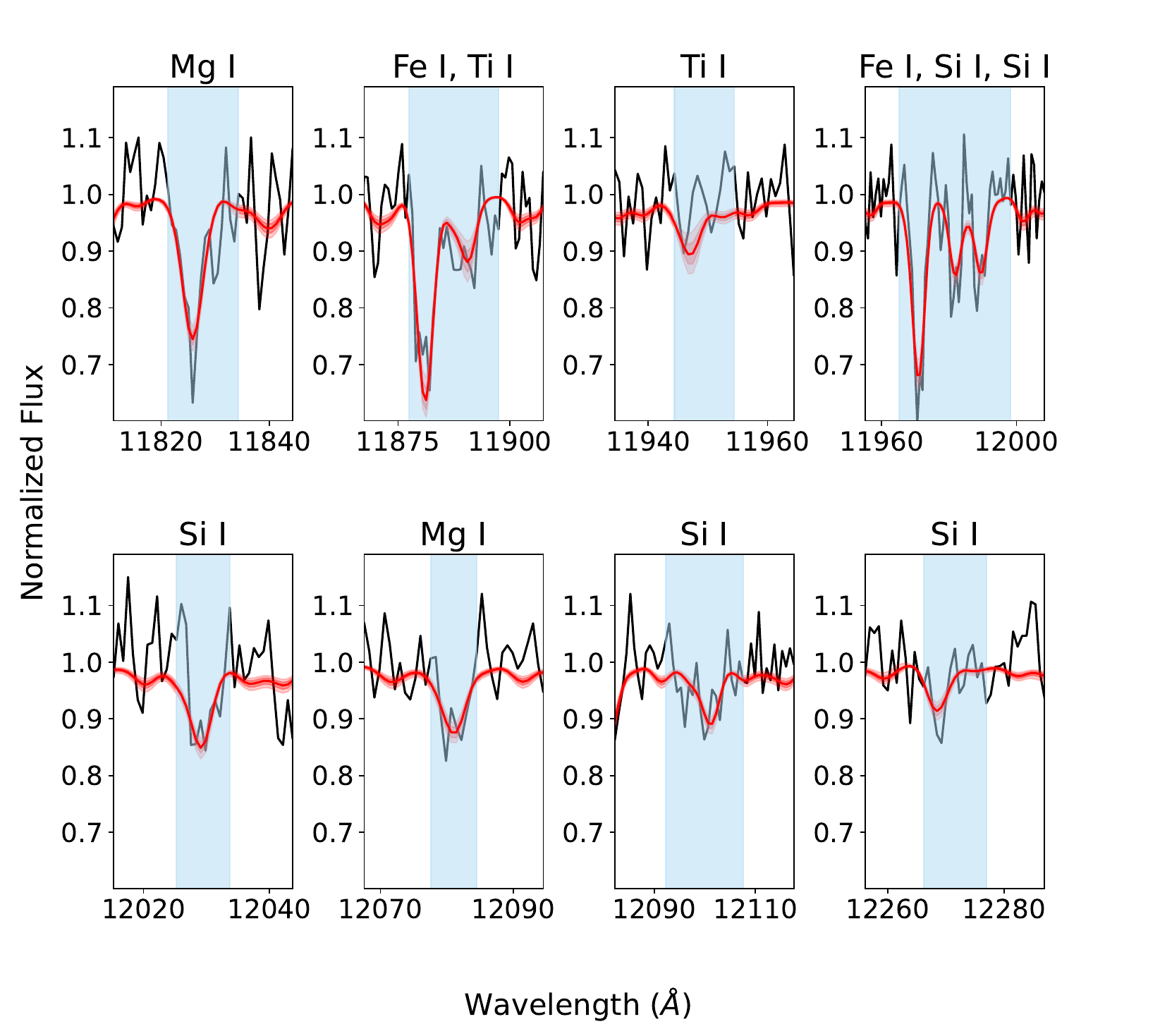}
    \caption{Same as Fig.~\ref{fit_ngc6822-103}, but showing only the fitted spectral regions and the area around them. The contours represent the 68.3\% confidence level errors. The spectral lines in each window are identified at the top of each panel.}
    \label{plot_ngc6822-103}    
\end{figure}

\begin{figure}
    \centering
    \includegraphics[width=1.0\columnwidth]{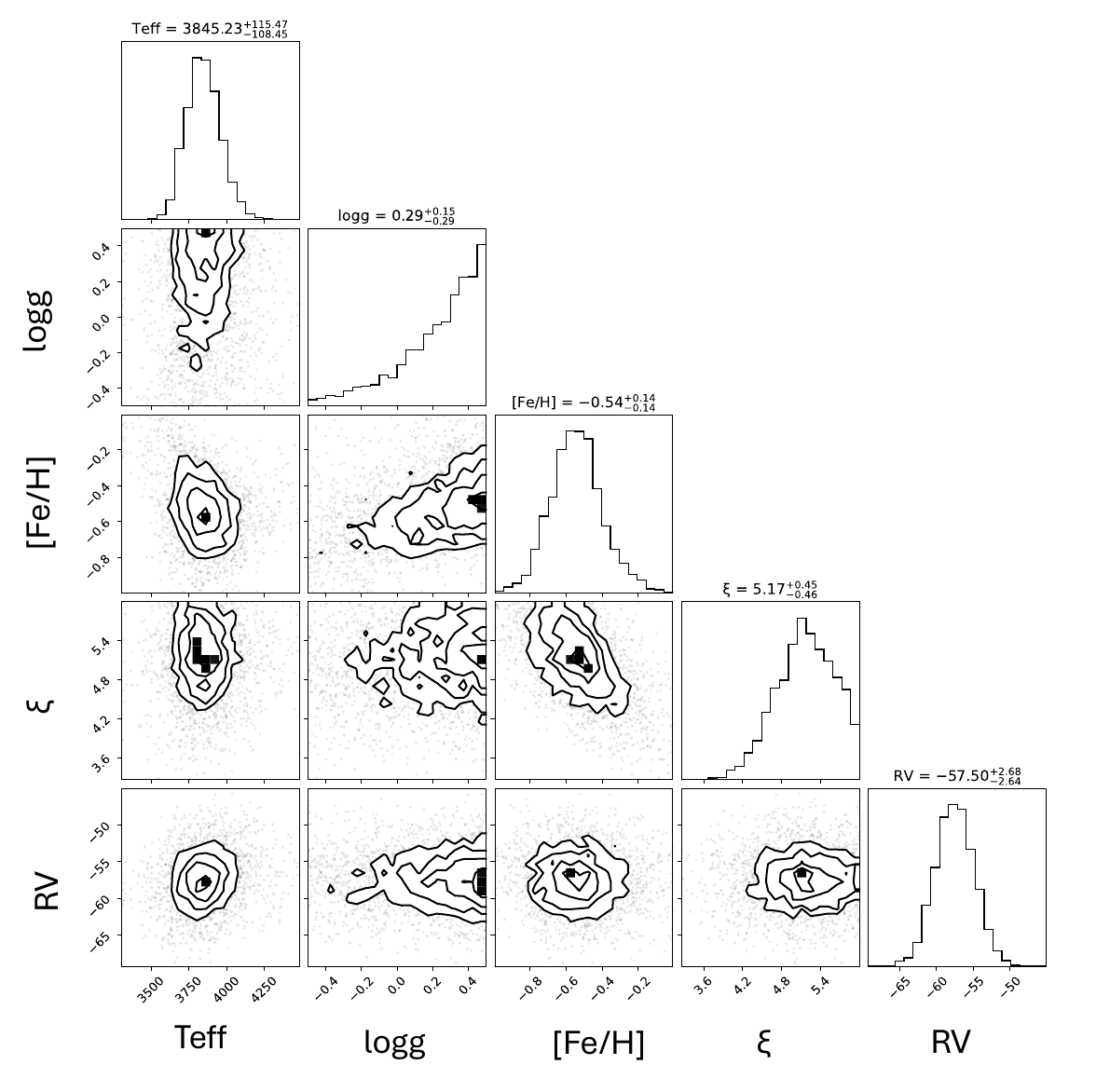}
    \caption{Cornerplot showing the samples from the posterior distribution of $T_{\rm eff}$, $\log$ $\it{g}$, [Fe/H], $\xi$, and radial velocity prior to the heliocentric correction (RV) for NGC6822-103. The solution vector is ($T_{\rm eff}$, $\log$ $\it{g}$, [Fe/H], $\xi$, RV)=(3845.23 K, 0.29, $-$0.54, 5.17 km~s$^{-1}$,$-$57.50 km~s$^{-1}$). Unfortunately, $\log$ $\it{g}$ is not well constrained. Contours represent regions of highest posterior probability, and their shape highlights potential correlations.}
    \label{corn_NGC6822-103}    
\end{figure}

Table~\ref{tab:fit_par} presents the parameters and their uncertainties obtained from the fitting process (described in Sect.~\ref{sec:fit_process}) for each object. It includes the $J$-band temperature, surface gravity ($\log g$), metallicity ([Fe/H]), microturbulence ($\xi$), and the radial velocity ($v_{\rm rad}$). We note that the accuracy of our metallicity values is comparable with \citet{Patrick2015}, despite our significantly lower S/N. Fig.~\ref{fit_ngc6822-103} shows the fit for NGC6822-103, demonstrating the quality of our spectral fitting. Fig.~\ref{plot_ngc6822-103} presents a zoom-in to the spectral windows used for the fitting process and the region used for the normalization around them. Fig.~\ref{corn_NGC6822-103} displays the cornerplot of the samples from the posterior distributions of the fitted parameters. The diagonal features the marginal posterior distributions, whereas the off-diagonal positions present scatter plots. Each scatter plot illustrates all samples of the column variable plotted against all samples of the row variable, highlighting potential correlations (e.g., RV is not correlated with any other parameter). For the rest of our targets, the best fit models are presented in Appendix~\ref{app:fits}. 

We compare the parameter values we derived with previous ones available for three of our sources in the literature. For the two RSGs in common with \citet{Patrick2015}, namely NGC6822-55 and NGC6822-175, we find values that agree within (our large) errors. We note that we were unable to include all the spectral features utilized by \citet{Patrick2015} due to the presence of artifacts or noise (see Figs. \ref{fit_NGC6822-55} and \ref{fit_NGC6822-175}). Specifically, we did not include Fe~\textsc{I} $\lambda$$\lambda$ 11882.847 and Ti~\textsc{I} $\lambda$$\lambda$ 11892.878 in the case of NGC6822-55 and Si~\textsc{I} $\lambda$ 12103.54 and Mg~\textsc{I} $\lambda$ 12083.346 in the case of NGC6822-175. We tested the impact of the additional removal of both Ti~\textsc{i} 11949.542 \AA\ and Si~\textsc{i} 12031.50 \AA\ for the case of NGC6822-55 because of their noisy profiles (Fig.~\ref{plot_NGC6822-55}) and found [Fe/H]$=-$0.29, a value that lies outside of our error bars (see Table~\ref{tab:fit_par}). All the other parameters varied within the error bars. The metallicity value reported for this object by \citet{Patrick2015} is $-$0.50 (same as our reported value, see Table~\ref{tab:fit_par}), leading us to the conclusion that these lines significantly impact metallicity; therefore, we cannot exclude them from the fit. 

\citet{Britavskiy2019} report a $T_{\rm eff}$ value of 3850~$\pm$ 80~K for WLM~14, estimated using SED fitting. Our calculated $T_{\rm eff}$ value agrees with it within 3--$\sigma$. However, we must emphasize that we could not incorporate all spectral lines into our fit because of artifacts or noise contaminating the line profiles (Fig.~\ref{fit_WLM_14}). We argue that achieving a better agreement than 3--$\sigma$ was not possible due to the low S/N of all of our spectra. Our $\log$ $\it{g}$ values have uncertainties that span almost the whole grid; therefore, we cannot constrain them. This problem is also present in \citet[][see their Table 4]{Patrick2015}. 
Furthermore, our estimated error bars for the stellar parameters of the NGC~6822 targets are comparable to those reported by \citet[][see their Table 2]{Chun2022}, despite their higher S/N spectra. We attribute this to weaker lines due to lower metallicity in their spectra, suggesting that high S/N at lower metallicity and low S/N at higher metallicity can yield similarly accurate stellar parameters.

We find radial velocity discrepancies between our measurements and previous values in the literature for NGC6822-55 and NGC6822-175. Specifically, for NGC6822-55 and NGC6822-175 we determine radial velocity values of $\sim$ $-$43$\pm$4 km~s$^{-1}$, while \citet{Patrick2015} report values of $-$67.9$\pm$3.1 km~s$^{-1}$ and $-$79.9$\pm$3.7 km~s$^{-1}$, respectively, resulting in a discrepancy of 30--40 km~s$^{-1}$. There are physical phenomena that could be responsible for such a discrepancy. Specifically, atmospheric convection causing stochastic shifts of the spectral lines can account for variations of $\sim$ 10 km~s$^{-1}$ \citep{Kravchenko2019, Kravchenko2021}, while large convective motions can explain velocity variations of $\sim$ 20 km~s$^{-1}$, a conclusion supported both from 3D simulations \citep[e.g.,][]{Kravchenko2021, Goldberg2022}, as well as spectroscopic and interferometric observations \citep[e.g.,][]{Josselin2007, Ohnaka2017}. Since we detected both spectroscopic and photometric variability (Table~\ref{tab:tar_prop} and Table~\ref{tab:p_l_short}, respectively) for both targets, indicating stellar activity on their surface, potentially in the form of convection, we argue that these phenomena are contributing to the discrepancy. We note that the lack of radial velocity standards during our observations and the exclusion of some spectral lines may also affect our radial velocity estimations. 

Finally, we compared our radial velocities with the values calculated by \citet{deWit2025}. We found inconsistent values for all of our common objects for which we could obtain $v_{\rm rad}$ measurements, namely NGC6822-55, NGC6822-175, NGC6822-70, and NGC6822-103, with the discrepancy ranging from $\sim$ 20 -- 180 km~s$^{-1}$. We argue that the main explanation for this disagreement is the inaccurate radial velocity calibration of the OSIRIS multi-object observations. This is due to the reduced wavelength range for the slits near the edges of the detector and the rotator angle, inducing shifts and therefore systematic errors up to 60 km~s$^{-1}$. Finally, the difference in resolution (i.e., $\sim$ 3200 for EMIR vs. $\sim$ 500--700 of OSIRIS) could contribute to the discrepancy.

The radial velocity we determine for WLM~14 is consistent with the value reported in the literature for this object \citep{Britavskiy2015}. The object is located in the southern part of WLM \citep[see Fig. 4 of][]{Britavskiy2015}, where such $v_{\rm rad}$ values are expected \citep[see Table 1 of][]{Bresolin2006}.

Previous studies in this regime have claimed that this type of analysis can be performed for spectra with S/N $\geq$ 100 \citep[e.g.,][]{Gazak2014, Lardo2015}. However, \citet{Patrick2017} and \citet{Chun2022} analyzed spectra with S/N $\geq$ 25 and S/N $\geq$ 40, respectively, producing reliable results. Since our error bars become significant below S/N $\sim$ 15, we conclude that the minimum S/N needed is 15.

\subsection{Light curve results} \label{sec:phot_res}

The L-S periodogram analysis yielded significant periods for 3 sources, namely NGC6822-52, IC10-26089, and WLM~14. The periodograms are presented in Appendix~\ref{app:periodograms}. The periods are within the expected range for RSGs \citep[i.e. 300--1000 days, see][]{Chatys2019}. In the case of IC10-26089, only the ZTF-$r$ band was usable. The periodograms of the other objects either revealed nonperiodic behavior or the signal was too noisy; therefore, no periods can be reported. Next, we calculated the expected periods of our sources based on the $K_{s}$-band P--L relations of \citet[][determined in the SMC]{Yang2012} and \citet[][determined for Galactic RSGs]{Chatys2019}, which have been found to be universal \citep[][i.e., their slopes are in agreement within uncertainties]{Ren2019}. The photometry was taken from Table~\ref{tab:tar_prop} and the distances from Table~\ref{tab:gal_prop}. Table~\ref{tab:p_l_short} contains all the properties determined using the light curves: the absolute $K_{s}$ magnitudes ($M_{K_{s}}$), the periods measured using the L-S periodograms, the expected periods calculated using the $K_s$-band P--L relations, the luminosity, the amplitude in $r_{\rm ZTF}$ and $W1$ and the mass-loss rates. The luminosity is taken from \citet{deWit2025}, who measured it using SED fitting. The mass-loss rates for the NGC~6822 targets are taken from \cite{Antoniadis2025}. We note that the absolute magnitude for IC10-26089 ($M_{K_{s}}=-$10.22 mag) is inconsistent with its high luminosity ($\log L / \lsun$ = 5.25), and the P--L periods are shorter than expected. The 2MASS photometry was obtained in 1997-2001, a time period for which we have no light curve coverage; therefore, we have no knowledge of how the variability of the object could have affected the measurement. However, the SED fit from \citet{deWit2025} is trustworthy.

We find two of our targets in NGC~6822 to display unusual variability ($\Delta r_{\rm ZTF}$ > 1 mag, as variability of $\sim$ 1 mag is typical for RSGs, \citealt{Levesque2017}), namely NGC6822-52 and  NGC6822-175 (see Table~\ref{tab:p_l_short} and Fig.~\ref{lc_ngc6822-52}, Fig.~\ref{lc_ngc6822-175}, respectively). Both the $\Delta r_{\rm ZTF}$ of 2.5 mag and $\Delta{\rm W1}$ of 0.5 mag of NGC6822-52 suggest an unusually strong pulsation. Such pulsations may occur during the last 10$^{4}$ yrs of their lifetime  \citep{Heger1997}. Currently, there are not enough available data to explore this possibility for NGC6822-52. In the case of NGC6822-175, its variability could be attributed to such a pulsation as well, despite its lower luminosity \citep{deWit2025}. 

\begin{figure*}
    \centering    \includegraphics[width=1.5\columnwidth]{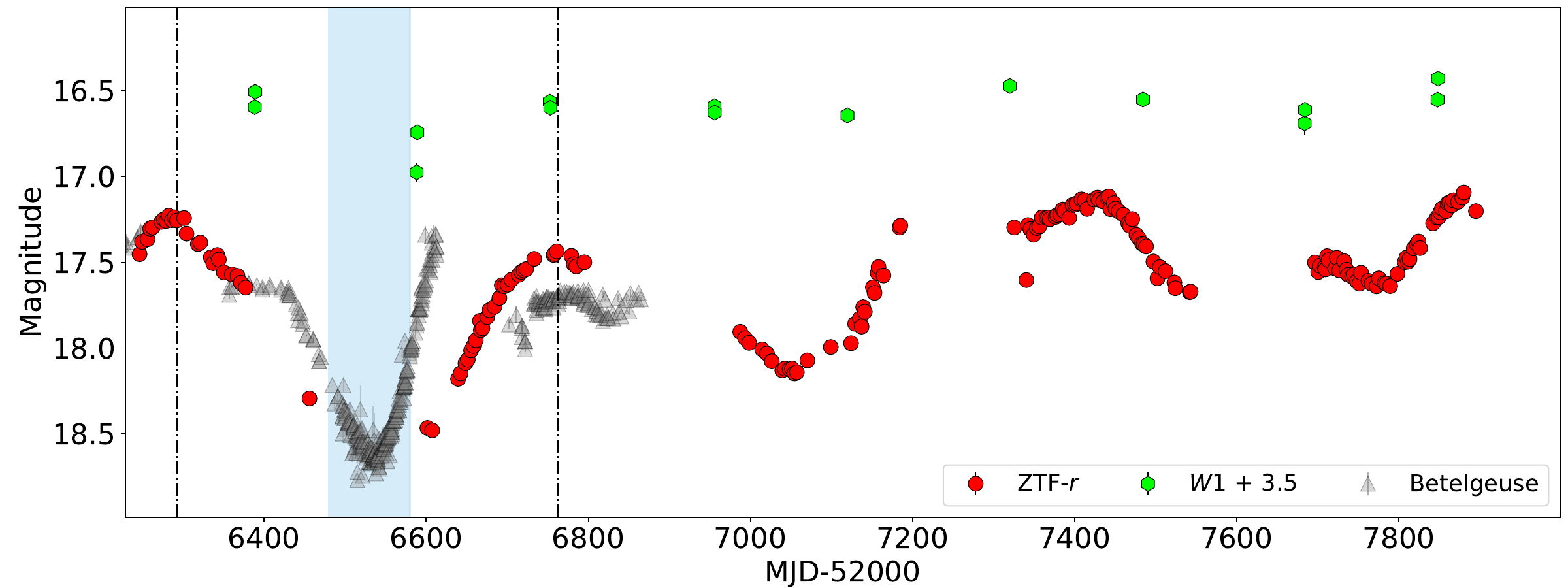}
    \caption{Zoom-in on the candidate-dimming event (marked with the dashed dotted lines) in the $r_{\rm ZTF}$ and $W1$-band light curves of NGC6822-175. Possible locations of the minimum are highlighted with the shaded region. The dimming event of Betelgeuse is shown for comparison (gray triangles). Data were retrieved from the AAVSO database.} 
    \label{dim_event_ngc6822-175}    
\end{figure*}

We searched the light curves for dimming events. These outstanding events display variability greater than 1 magnitude and have a bell-like shape \citep[see e.g., Fig. 4 of][]{Munoz-Sanchez2024}. We found a candidate-dimming event in NGC6822-175, with an estimated duration of 550 days (between MJD 58292 -- 58762) and an amplitude exceeding 1.1 mag in the $r_{\rm ZTF}$ light curve. The exact amplitude and nature are uncertain because the minimum lacks photometric coverage. ZTF only covers the decrease in magnitude and the recovery. This event is also present in the \textit{WISE} $W1$ light curve, with an amplitude of $\sim$ 0.5 mag, as there is a local minimum during the time span of the event. Fig.~\ref{dim_event_ngc6822-175} compares the candidate-dimming event in $r_{\rm ZTF}$ to the AAVSO photometry of the Great Dimming of Betelgeuse and overplots the $W1$ light curve. We discuss the properties of this candidate-dimming event in Sect.~\ref{cdiming_prop}. We note that the semi-periodic variability in the optical light curve in the $\sim$ 1000 days preceding the event displayed minima with increasing depth, while after the dimming, the minima became shallower. The same behavior is observed in the infrared light curve (Fig.~\ref{lc_ngc6822-175}).

\begin{table*}[h]
\small
\centering
\caption{Effective temperatures for our targets obtained using different methods.}
\label{tab:teff}
\begin{tabular}{l l l l l l l l}
\hline
\hline
        Object & $T^{J,\, \rm fit}_{\rm eff}$ (K) & $T^{J,\, \rm P15}_{\rm eff}$ (K)\tablefootmark{a} & $T^{J-K_{s}}_{\rm eff}$ (K)\tablefootmark{b} & $T^{\rm MARCS}_{\rm eff}$ (K)\tablefootmark{c} & $T^{\rm TiO}_{\rm eff}$ (K) & $T^{scaled}_{\rm eff,J}$ (K)\tablefootmark{f} & $T^{J-K}_{\rm eff}$ (K) \tablefootmark{g}  \\
        \hline
        IC10-26089 & - & - & 4045 $\pm$ 48 & 4058 $\pm$ 125 & 4273\tablefootmark{d}$\pm$50& 4137$\pm$12 & 3800 $\pm$ 150 \\
         NGC6822-52 & - & - & 4027 $\pm$ 48 & 4029 $\pm$ 124  & 3351\tablefootmark{d}$\pm$50& -& 3802 $\pm$ 150   \\
         NGC6822-55 & 4286$^{+136}_{-145}$  & 3860$\pm$70  & 4034 $\pm$ 48 & 4044 $\pm$ 124  & 4021\tablefootmark{d}$\pm$50 & 4075$\pm$12 & -    \\
         NGC6822-70 & 3688$^{+153}_{-124}$  & - & 3957 $\pm$ 48 & 3848 $\pm$ 124  & 3520\tablefootmark{d}$\pm$50 & 3955$\pm$12 & 3631 $\pm$ 150\\
         NGC6822-103 & 3845$^{+115}_{-108}$ & - & 3974 $\pm$ 48 & 3892 $\pm$ 124  & 3681\tablefootmark{d}$\pm$50  & 3993$\pm$12 & 3715 $\pm$ 150  \\
         NGC6822-175 & 4383$^{+80}_{-108}$ & 3985$\pm$70   &  4003 $\pm$ 48 & 3964 $\pm$ 124  & 3537\tablefootmark{d}$\pm$50 & 3959$\pm$12 & 3548 $\pm$ 150    \\
         WLM~14 & 4349$^{+104}_{-233}$ & - & 4142 $\pm$ 48 & 4338 $\pm$ 122  & 3850\tablefootmark{e}$\pm$80 & 4034$\pm$12 & - \\   \hline         
\end{tabular}
\tablebib{
    $^{(a)}$ Mean value of the two $T_{\rm{eff}}$ reported in \citet{Patrick2015}. Error bar reported here is the mean of their associated error bars. \\
    $^{(b)}$ Calculated using the empirical $T_{\rm eff}(J-K_{s})_{0}$ relation in \citet[Table 5;][]{deWit2024}. Error bars determined using error propagation. \\
    $^{(c)}$ Calculated using the theoretical $T_{\rm eff}(J-K_{s})_{0}$ relation from \textsc{marcs} synthetic photometry in \citet[Table 5;][]{deWit2024}. Error bars determined using error propagation.\\
    $^{(d)}$ Values taken from \citet{deWit2025}. \\
    $^{(e)}$ Value taken from \citet{Britavskiy2019}. \\
    $^{(f)}$ Values calculated using the scaling relation for the $J$-band of \citet[Eq. 2;][]{deWit2024}.\\
    $^{(g)}$ Values taken from \citet{Dimitrova2020} for IC10-26089 and \citet{Dimitrova2022} for the targets in NGC~6822.}

\end{table*}

\begin{table*}[h]
\small
\centering
\caption{Properties of the light curves} \label{tab:p_l_short}
\begin{tabular}{l r c r r l c l l}
\hline
\hline
ID & $M_{K_{s}}$ & P & $P_{\rm{Yang}}$$^{1}$ & $P_{\rm{Chatys}}$$^{2}$ & \small{log(L$^{\rm{SED}}$/\lsun)} & $\Delta r_{\rm{ZTF}}$& $\Delta W1$ & $\dot{M}$$^{5}$ \\
& (mag) & (days) & (days) & (days) & (dex) & (mag) & (mag) & ( $\times$ 10$^{-6}$ \msun~yr$^{-1}$) \\
\hline
IC10-26089 & $-$10.22 & 482 & 406 & 395  & 5.25$^{+0.04}_{-0.04}$ $^{3}$ & 0.92 $\pm$ 0.07 & 0.45 $\pm$ 0.03 & -\\
NGC6822-52 & $-$10.94 & 693 & 673 & 904  & 5.10$^{+0.06}_{-0.05}$ $^{3}$ & 2.49 $\pm$ 0.03 & 0.50 $\pm$ 0.03 & 3.98$^{+1.38}_{-1.49}$ \\
NGC6822-55 & $-$11.13 & - & 769 & 1112  & 5.21$^{+0.05}_{-0.05}$ $^{3}$ & 0.77 $\pm$ 0.02 & 0.27 $\pm$ 0.02 &0.18$^{+0.24}_{-1.78}$  \\
NGC6822-70 & $-$10.73 & - & 581 & 715  &5.01$^{+0.05}_{-0.05}$ $^{3}$ & 0.72 $\pm$ 0.03 & 0.27 $\pm$ 0.03 & 2.01$^{+1.70}_{-0.79}$  \\ 
NGC6822-103 & $-$10.70 & - & 569 & 691  &5.01$^{+0.05}_{-0.05}$ $^{3}$ & 0.58 $\pm$ 0.03 & 0.21 $\pm$ 0.03 & 12.18$^{+0.60}_{-0.77}$  \\
NGC6822-175 & $-$10.11 & - & 376 & 346  & 4.80$^{+0.05}_{-0.05}$ $^{3}$ & > 1.07 $\pm$ 0.04 & 0.47 $\pm$ 0.04 & 2.05$^{+7.26}_{-1.81}$ \\ 
WLM~14 & $-$10.46 & 463 & 482 & 528  & 4.87$^{+0.1}_{-0.1}$ $^{4}$ & 0.65 $\pm$ 0.06 & 0.25 $\pm$ 0.07 & - \\
\hline
\end{tabular}
\tablebib{
    (1) Calculated using the P-L relation of \citet{Yang2012};
    (2) Calculated using the P-L relation of \citet{Chatys2019};
    (3) \citet{deWit2025} ;
    (4) \citet{Britavskiy2019} ;
    (5) \citet{Antoniadis2025}
    }
\end{table*}

\subsection{Evidence for episodic mass loss}\label{dis_m_loss}
Episodic mass loss can be traced by taking into account both the photometric and spectroscopic variability of our sources. Specifically, we search for nontypical photometric and spectral variability. Also, we calculate the median absolute deviation in $W1$ (MAD$_{\rm W1}$) and compare it to the right panel of Fig.~12 of \citet{Yang2023}. If the object lies in the region after the "kink" feature where mass-loss rates increase, it is an indication of strong mid-IR variability, meaning that either it has experienced or will experience episodic mass loss. Thus, we identify four targets out of seven that exhibit evidence for episodic mass loss:

\begin{itemize}
\item{NGC6822-52 has a high luminosity (log(L${^{\rm{SED}}}$/\lsun)= 5.10) and a long period (693 days). It is the most photometrically variable star ($\Delta r_{\rm ZTF}$ $\sim$ 2.5 mag and $\Delta W1$ $\sim$ 0.5 mag) in the sample (Fig.~\ref{lc_ngc6822-52}). It also displays spectroscopic variability transitioning from a spectral type of M1 to M4 within $\sim$ 40 years (see Table~\ref{tab:tar_prop}). It has a MAD$_{\rm W1}$ value of 0.10 mag and, given its luminosity, it is situated after the kink feature. Considering all the information presented above and its IR excess, we conclude that it is a strong candidate for episodic mass loss.}
\\
\item{NGC6822-55 is one of the most luminous (log(L$^{\rm{SED}}$/\lsun)= 5.21) sources in our sample. Its light curve displays significant variability ($\Delta r_{\rm ZTF}$ $\sim$ 0.8 mag and $\Delta W1$ $\sim$ 0.3 mag; Fig.~\ref{lc_ngc6822-55}). Its spectral type has changed from M0 to K7 between 2008 and 2020 (see Table~\ref{tab:tar_prop}), indicating that we have a K-type RSG with IR excess. The mid-IR photometry was obtained at a different epoch, indicating that the object was in a cooler, more extended state and therefore losing mass at a higher rate. This caused its colors to be redder due to the formation of dust. It has since returned to a hotter, more compact state observed by spectroscopy \citep{deWit2024}. Furthermore, \citet{deWit2025} report that given its low mass-loss rate (see Table~\ref{tab:p_l_short}), an eruptive mass-loss event should have taken place to produce the increased amounts of dust measured by the mid-IR photometry. Its MAD$_{\rm W1}$ value is 0.05 mag; therefore, it is located clearly after the kink feature, favoring the episodic mass loss scenario. The $J$-band $T_{\rm{eff}}$ derived in this work is higher than the one estimated by \citet[][spectrum obtained in 2013]{Patrick2015}, confirming this temperature rise.} 
\\
\item{IC10-26089 is the most luminous source in our sample (log(L$^{\rm{SED}}$/\lsun)= 5.25), yet it has a small period (482 days). It has a K4 spectral type; however, its position in the CMD (Fig.~\ref{fig:cmd}) suggests that it is a dusty star. This behavior is similar to NGC6822-55, indicating that we have identified another candidate for episodic mass loss. Once again, its MAD$_{\rm W1}$ value of 0.06 mag means that it can be found after the kink feature, favoring the scenario of a past eruptive mass-loss event. However, we do not have a mass-loss rate estimation, and there is no other optical spectrum in the literature for this target. Also, we were unable to model its $J$-band spectrum. It displays strong photometric variability ($\Delta r_{\rm ZTF}$ $\sim$ 0.9 mag and $\Delta W1$ $\sim$ 0.5 mag; Fig.~\ref{lc_ic1026089}). }
\\
\item{NGC6822-175 has the lowest luminosity (log(L$^{\rm{SED}}$/\lsun)= 4.80) of the sample and the smallest period (376 days). It has a candidate-dimming event in its light curve ($\Delta r_{\rm ZTF}$ > 1.1 mag and $\Delta W1$ $\sim$ 0.5 mag; Fig.~\ref{lc_ngc6822-175}), a typical indication of episodic mass loss. Its mass-loss rate is similar to NGC6822-70. It also experiences spectral variability, transitioning from M1 to M4 between 2008 and 2020.
Along with its luminosity, its MAD$_{\rm W1}$ value of 0.09 mag places it after the kink feature. Therefore, the star displays strong evidence of at least one significant past mass ejection.}
\end{itemize}

NGC6822-70 exhibits moderate photometric variability ($\Delta r_{\rm ZTF}$ $\sim$ 0.7 mag and $\Delta W1$ $\sim$ 0.3 mag; Fig.~\ref{lc_ngc6822-70}); however, there is no other optical spectrum in the literature. Similarly, NGC6822-103, the target with the highest mass-loss rate (see Table~\ref{tab:p_l_short}), displays mild photometric variability ($\Delta r_{\rm ZTF}$ $\sim$ 0.6 mag and $\Delta W1$ $\sim$ 0.2 mag; Fig.~\ref{lc_ngc6822-103}) as well but there is no other evidence for episodic mass loss. Also, the position of WLM~14 in the CMD does not suggest a dusty star. It displays no spectral variability; however, its photometric variability is also modest ($\Delta r_{\rm ZTF}$ $\sim$ 0.7 mag and $\Delta W1$ $\sim$ 0.3 mag; Fig.~\ref{lc_wlm_14}). Currently, there is no mass-loss rate estimation for this target. However, this could be potentially revisited in the future using \textit{James Webb Space Telescope (JWST)} data now available for WLM. 

To determine whether moderate photometric variability is a sufficient clue for episodic mass loss, we calculated the MAD$_{\rm W1}$ for NGC6822-103, NGC6822-70 and WLM~14 and found values of 0.05, 0.04 and 0.06 mag, respectively, meaning that the objects lie in the region after the kink feature therefore they could experience episodic mass loss in the future or, if they have already experienced such events, that we just do not have the data to trace them. 

\section{Discussion} \label{secDis}

\subsection{Hertzsprung–Russell diagram}\label{sec:HR}

\begin{figure}
    \centering    \includegraphics[width=1.0\columnwidth]{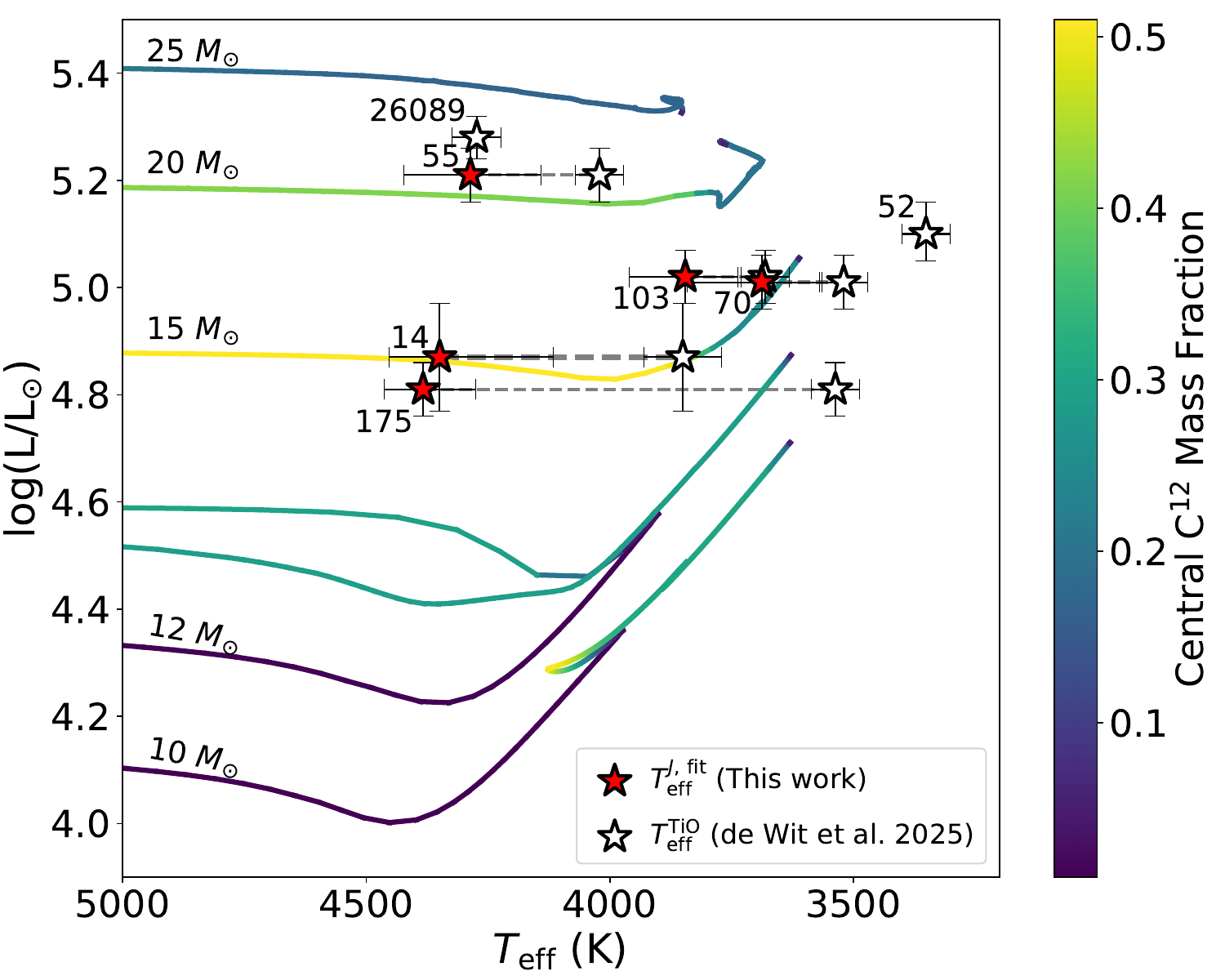}
    \caption{HR diagram for targets with luminosity estimates. Red stars denote the $T^{J,\, \rm fit}_{\rm eff}$ from this work and white stars denote the $T^{\rm TiO}_{\rm eff}$ from \citet{deWit2025}. Each red star is labeled with the ID number of the corresponding target. Dashed lines connect measurements for the same target.}
    \label{hr_diagram}    
\end{figure}

We present a Hertzsprung–Russell (HR) diagram, where we compare our sample to evolutionary tracks and search for outliers (Fig.~\ref{hr_diagram}). We include \textsc{mist} stellar tracks \citep{Dotter2016, Choi2016} for rotating stars ($v/$$v_{\rm{crit}}$ = 0.4) of 10, 15, 20 and 25 $\msun$ and [Fe/H]=$-$0.50 to compare the evolutionary predictions to the evolutionary status of our targets, given by their derived $T_{\rm{eff}}$ and $\log L$. We choose the \textsc{mist} tracks over the Geneva ones because the former provide a better match to the high-mass RSG population of the LMC \citep{Yang2021, deWit2023}. As most of our targets reside in NGC~6822, we use tracks consistent with its average metallicity value of [Z]=$-$0.52 reported by \citet{Patrick2015}. We assume that [Z]=[Fe/H], following the assumptions made in the same work. For each source, we show both the $T^{\rm TiO}_{\rm eff}$ (empty stars) from \citet{deWit2025} and $T^{J,\, \rm fit}_{\rm eff}$ (red stars) calculated in this work. For NGC6822-52 and IC10-26089 we only present the $T^{\rm TiO}_{\rm eff}$. 

We limit our HR diagram to the region occupied by RSGs. Specifically, \citet{Antoniadis2025} have determined a luminosity range between 3.2 and 5.26 for their candidate RSGs in NGC~6822. However, it is unclear whether objects with $\log$(L/\lsun) < 4 are in fact RSGs or asymptotic giant branch stars \citep[see e.g.,][]{Yang2024}. Therefore, we adopt the widely accepted lower limit of $\log$(L/\lsun) = 4 for RSGs. Furthermore, there is some controversy about the upper luminosity limit of RSGs \citep[][]{Humphreys1979, Davies2018, McDonald2022, Munoz-SanchezWOHG642024, Maravelias2025}. We adopted an upper limit of $\log$(L/\lsun) = 5.5.
 
We note that NGC6822-70 and NGC6822-175 are close to the Hayashi limit when considering their $T^{\rm TiO}_{\rm eff}$. However, when we take into account their $T^{J,\, \rm fit}_{\rm eff}$, they are brought back to the permitted region of the HR diagram. We attribute this behavior to the fact that the TiO-band method provides cooler $T_{\rm{eff}}$ estimates. We remark that in the case of NGC6822-175 the discrepancy between the $T^{J,\, \rm fit}_{\rm eff}$ and the $T^{\rm TiO}_{\rm eff}$ cannot be explained by the shortcomings of the TiO-band method alone \citep[which are responsible for differences of 100--200~K, see][]{Davies2013}. We do not find any of our targets to be outliers in the HR diagram, especially when their $J$-band temperatures are taken into account.

We compare with the HR diagram constructed by \citet[][their Fig.~11]{Patrick2015} and find that most of our RSGs occupy the same luminosity regime ($\log$ L $\geq$ 4.8) but most of them are not in the same $T_{\rm eff}$ regime, as theirs are at $T_{\rm eff}$ $\leq$ 4000~K. Since we only have two targets in common with them and we achieved agreement within 3 $\sigma$ with their derived $T_{\rm eff}$ (see previous sections) and our error bars are significantly larger than theirs (see Table~\ref{tab:teff}), we refrain from drawing strong conclusions. 

However, there are pitfalls to the evolutionary tracks.
Specifically, \textsc{mist} stellar tracks use a fixed value of 1.25 for the mixing length  \citep[$\alpha_{MT}$,][]{Choi2016}; however, recent studies have shown that fixing the $\alpha_{MT}$ value is not optimal for all kinds of stars \citep[e.g.,][]{Bonaca2012, Tayar2017}. Furthermore, it is known that the mixing length has a metallicity dependence both for high and for low mass stars \citep{Chun2018, Joyce2018}, namely, it increases with metallicity \citep[e.g.,][]{Bonaca2012}. Furthermore, the \textsc{mist} stellar tracks are calculated using 1D stellar atmosphere models; thus, they cannot describe 3D phenomena such as mass loss or convection \citep[][]{Choi2016}, in turn failing to accurately predict the evolution of targets such as NGC6822-55 and NGC6822-175 (see previous sections).

We observe that $T^{J,\, \rm fit}_{\rm eff}$ are not near the Hayashi limit, as opposed to the $T^{\rm TiO}_{\rm eff}$. This is a known limitation of the $T^{J,\, \rm fit}_{\rm eff}$ values \citep[see Fig.~9 of][]{Patrick2015} that applies to the $T^{scaled}_{\rm eff,J}$ as well \citep[][see the left panel of their Fig.~14]{deWit2024}. However, the Hayashi limit shifts to higher temperatures for lower metallicities and higher mixing length values. \citet{Chun2022} has shown that for low metallicity, a mixing length value of $\sim$ 2--2.5 shifts the Hayashi limit to the $T^{J,\, \rm fit}_{\rm eff}$ locus (see their Figs.~5 and 6). Therefore, we speculate that stellar tracks with a $\alpha_{MT}$ value within this range could place the Hayashi limit closer to our  $T^{J,\, \rm fit}_{\rm eff}$ values.

\vspace{0.6cm}
\subsection{Effective temperatures} \label{dis_teff}

We compare the effective temperatures obtained for our targets using the various methods described above, searching for overall trends. The $T^{TiO}_{\rm eff}$ temperatures obtained from modeling the TiO bands are lower than the $T^{J,\, \rm fit}_{\rm eff}$ obtained by fitting the $J$-band. This is in agreement with the conclusions of \cite{Davies2013} and is also valid for our derived $T^{scaled}_{\rm eff,J}$. We note that for the four targets we have in common with \citet{Dimitrova2022}, namely NGC6822-52, NGC6822-70, NGC6822-103 and NGC6822-175, the $T_{\rm eff}$ values determined in that work ($T^{J-K}_{\rm eff}$) are only hotter than $T^{TiO}_{\rm eff}$. In contrast, in the case of IC10-26089, which was also studied by \cite{Dimitrova2020}, their $T^{J-K}_{\rm eff}$ value is the lowest one measured for the object. Furthermore, $T^{J-K_{s}}_{\rm eff}$ and $T^{\rm MARCS}_{\rm eff}$ are in agreement within the error bars for all of our objects. For most of our targets, both $T^{J,\, \rm fit}_{\rm eff}$ and $T^{scaled}_{\rm eff,J}$ are even hotter than $T^{J-K_{s}}_{\rm eff}$. Notable exceptions are NGC6822-103 and NGC6822-70 where $T^{J-K_{s}}_{\rm eff}$ values are hotter than $T^{J,\, \rm fit}_{\rm eff}$ but cooler than $T^{scaled}_{\rm eff,J}$. In the case of NGC6822-55 $T^{J-K_{s}}_{\rm eff}$, $T^{\rm MARCS}_{\rm eff}$, $T^{TiO}_{\rm eff}$ and $T^{scaled}_{\rm eff,J}$ agree within the error bars, whereas for NGC6822-175 $T^{J,\, \rm P15}_{\rm eff}$, $T^{J-K_{s}}_{\rm eff}$ and $T^{\rm MARCS}_{\rm eff}$ are in agreement within the error bars. Finally, we highlight that in the respective cases of NGC6822-103 and NGC6822-175, the values of our error bars are comparable with the ones obtained by \citet{Patrick2015}.

We note the case of WLM~14 where the effective temperatures obtained have a $\sim$ 500~K span. The relation used to calculate $T^{J-K_{s}}_{\rm eff}$(see Table~\ref{tab:teff}) was calibrated on a sample of RSGs located in the Magellanic Clouds where the metallicity ranges from 0.2 to 0.79 \zsun\  \citep[see Table 5 of][for details]{deWit2024}, whereas the metallicity of WLM is lower (see Table~\ref{tab:gal_prop}); therefore, we extrapolated the relation to obtain this value. We are unable to use the scaling relation for NGC6822-52 because its $T^{TiO}_{\rm eff}$ value is below the $T_{\rm{eff}}$ range where the relation is applicable \citep[see Sect. 4.3 of][]{deWit2024}. Furthermore, this $T^{TiO}_{\rm eff}$ value is close to the lower edge of the grid of \citet{deWit2025}, so we urge caution in its interpretation.
As a final note, when we cross-matched our IC~10 target with the catalog of \citet{Dimitrova2020}, we found two sources with identical coordinates matching within a $\ang{;;1.0}$ radius. We selected the target with a \textit{Gaia} parallax to be the correct match, as our target had a \textit{Gaia} parallax.

\subsection{Metallicity} \label{sec:metal}
\citet[][]{Patrick2015} report that NGC~6822 has a low significance metallicity gradient within 1~kpc from the galactic center utilizing 11 RSGs. Specifically, as the distance from the galactic center increases, the metallicity decreases. They report an average metallicity of $-$0.52, while individual RSGs have values ranging from $-$0.20 to $-$0.78. This average value is consistent with the values reported by studies that utilized blue supergiants \citep{Muschielok1999, Venn2001}. 

We have modeled 4 targets in NGC~6822. For NGC6822-55, which is at the center of the galaxy, we find a metallicity value of $-$0.50. For NGC6822-70 and NGC6822-175, which are located in the outer part of the galaxy, we find $-$0.83 and $-$0.81, respectively. Given the objects' respective locations, these values are anticipated. However, NGC6822-103 has a metallicity value of $-$0.54, although it is situated on the outskirts of the galaxy. 

\subsection{Candidate-dimming event}\label{cdiming_prop}
We compare the properties of the proposed dimming event in NGC6822-175 to events that have occurred in other RSGs, and in particular Betelgeuse. We argue that the light curve of NGC6822-175 after the candidate-dimming event could be similar to that of Betelgeuse after its own dimming episode known as the "Great Dimming" \citep{Montarges2021, Dupree2022}, see \citet{Jadlovsky2024} for a recent light curve, but because of the limited coverage compared to Betelgeuse, we refrain from drawing strong conclusions. Studies of Betelgeuse after the Great Dimming have shown that it exhibits smaller amplitude variability, pulsating in its first overtone, which has a period of $\sim$ 200 d \citep[i.e., half of its period before the Great Dimming;][]{Granzer2022}. 
Betelgeuse is predicted to revert to its pre-Great Dimming behavior within the next 5-10 yrs \citep{MacLeod2023}. It is likely that our source could exhibit similar behavior. In the case of $\rm{[W60]}$~B90, dimming events are recurring \citep[with a periodicity of $\sim$ 11.8 yrs, see][]{Munoz-Sanchez2024}; however, for other RSGs such as Betelgeuse and RW Cep, the dimming events are unique on a timescale of 100 yrs. As we do not have a period for NGC6822-175 (see Table~\ref{tab:p_l_short}), we cannot compare directly with Betelgeuse or $\rm{[W60]}$~B90. NGC6822-175 ($\log L/ \lsun$ = 4.8, see Table~\ref{tab:p_l_short}) is less luminous than Betelgeuse and $\rm{[W60]}$~B90 ($\log L/ \lsun$ = 5.10 and 5.32, respectively; see Table 5 of \citealt{Munoz-Sanchez2024}). Our light curve coverage stops $\sim$ 5.5 yrs after the event, in which no other dimming event has been detected. This interval is likely too short for us to draw any meaningful conclusions or make reliable predictions regarding the source's behavior.

Recently, \citet{Munoz-Sanchez2024} suggested that the timescale of these events is connected to the radius of the RSGs that host them, as more extended atmospheres need more time to return to stability. Using the Stefan-Boltzmann law, we estimated the radius of NGC6822-175 to be within the range of 530--678 $\rsun$ depending on $T_{\rm eff}$ (see Table~\ref{tab:teff}), which is close to the lower end of the radii range reported for Betelgeuse  \citep[750--1000 $\rsun$,][]{Joyce2020, Kravchenko2021}. The rise time for NGC6822-175 after its event is $\sim$ 200--300 days, depending on the possible locations of the minimum (see Fig.~\ref{dim_event_ngc6822-175}), which is similar to Betelgeuse, but only around half the recovery time of $\rm{[W60]}$~B90, which has a significantly larger radius \citep[$\sim$ 1200 $\rsun$,][]{Munoz-Sanchez2024}. This result further supports the proposed connection between the radius of RSGs and the timescale of their dimming events.

\section{Summary and conclusions} \label{sec:summary}

We have conducted a comprehensive study of seven mass-losing RSGs in nearby low-metallicity galaxies. We obtained usable spectra for five out of seven targets with EMIR on GTC. We modeled these spectra using MARCS models corrected for nonlocal thermal equilibrium effects and obtained physical parameters for these RSGs. Additionally, we calculated $T_{\rm{eff}}$ values for all of our targets using empirical and scaling relations presented in \citet{deWit2024}, which are in agreement with the values determined by the modeling. We presented their optical, near-IR, and mid-IR light curves, discovering two RSGs (NGC6822-52 and NGC6822-175) with unusual photometric variability. The light curves yielded periods for three RSGs. We discovered a candidate-dimming event in NGC6822-175 with an estimated duration of 550 days and depth in $r$ over 1.1 mag. Monitoring of this object could reveal whether this phenomenon is recurring or if this was a unique event. We also improved the spectral classification from the available optical spectra, finding four sources exhibiting spectral variability.

Combining all of our results, we identified four RSGs (NGC6822-52, NGC6822-55, NGC6822-175, and IC10-26089) that are candidates for having undergone episodic mass loss. We suggest that NGC6822-70, NGC6822-103, and WLM~14 can be confirmed as candidates for episodic mass loss with additional spectroscopy. This work shows that both multi-epoch spectroscopy and time-series photometry are valuable for the characterization of episodic mass loss for dusty RSGs, as they trace the changes it causes to the stellar atmospheres. 

However, RSGs with episodic mass loss are rare \citep{deWit2024}. The development of machine--learning photometric classifiers \citep[e.g.,][]{Maravelias2022} enables studies of large samples of RSGs (over 100,000 RSGc in $\sim$20 nearby galaxies have been identified by \citet{Maravelias2025}) and the characterization of the frequency of episodic mass loss, shedding light on their mass-loss history. Future surveys such as the Rubin Observatory Legacy Survey of Space and Time (LSST) will provide light curves for many objects, and \textit{JWST} is providing mid-IR photometry, critical for collecting large samples of targets in distant galaxies. Along with spectroscopy, these data will be paramount to discovering RSGs with episodic mass loss and Levesque--Massey variables, as well as determining the physical mechanisms causing these phenomena.

\begin{acknowledgements}
EC, SdW, AZB, GMS, GM, and KA acknowledge funding support from the European Research Council (ERC) under the European Union’s Horizon 2020 research and innovation program ("ASSESS", Grant agreement No.
772086). M.M.R-D acknowledges financial support through Spanish grant PID2019-105552RB-C41(MCIU) and from Comunidad de Madrid through the Estimulo a la Excelencia para Profesores Universitarios Permanentes (EPDU-INV/2020/008). 
Based on observations made with the Gran Telescopio Canarias (GTC), installed at the Spanish Observatorio del Roque de los Muchachos of the Instituto de Astrofísica de Canarias, on the island of La Palma. This work is (partly) based on data obtained with the instrument EMIR, built by a Consortium led by the Instituto de Astrofísica de Canarias. EMIR was funded by GRANTECAN and the National Plan of Astronomy and Astrophysics of the Spanish Government. This work is based in part on observations made with the Spitzer Space Telescope, which is operated by the Jet Propulsion Laboratory, California Institute of Technology, under a contract with NASA. This work has made use of data from the European Space Agency (ESA) mission {\it Gaia} (\url{https://www.cosmos.esa.int/gaia}), processed by the {\it Gaia}
Data Processing and Analysis Consortium (DPAC, \url{https://www.cosmos.esa.int/web/gaia/dpac/consortium}). Funding for the DPAC has been provided by national institutions, in particular the institutions participating in the {\it Gaia} Multilateral Agreement. We acknowledge with thanks the variable star observations from the AAVSO International Database contributed by observers worldwide and used in this research.
We thank John Pritchard for his valuable assistance with the \textsc{Molecfit} installation and execution. We thank Lee Patrick for providing his catalog of RSG candidates and for useful discussions during various stages of the paper. We thank Patricia Whitelock and John Menzies for providing us with their $JHK_{s}$ photometric data.
We thank Despina Hatzidimitriou, Philip Wiseman, and Karan Dsilva for useful discussions.
\end{acknowledgements}

\bibliographystyle{aa} % style aa.bst
\bibliography{ref.bib}

\clearpage
\begin{appendix}
\onecolumn

\section{Observation log}\label{app:obs_log}
This Appendix provides the observing log of our long slit observations with EMIR at GTC.

\begin{table*}[h]
\small
\centering
\caption{Log of observations} \label{tab:log}
\begin{tabular}{l r r c r c c c r}
\hline
\hline
ID & R.A.  & Dec.  & UT Date & $t_{\rm{exp.}}$ & Airmass & Seeing & Moon & S/N  \\
             & \small{(\degr)} & \small{(\degr)} & & \small{(s)} & & ($\arcsec$) & &    \\
\hline
IC10-26089 & 5.02108 & 59.30108 & 2022 Sep 09 & 2280 & 1.30 & 0.8 & Bright & 6  \\
                      
NGC6822-52 & 296.21295 & $-$14.73218 & 2022 Oct 12 & 960 & 1.41 & 1.0 & Dark & 4 \\
              
NGC6822-55 & 296.23214 & $-$14.86554 & 2022 Oct 11 & 960 & 1.40 & 0.9 & Dark & 13 \\
      
NGC6822-70 & 296.22841 & $-$14.73003 & 2022 Oct 12 & 960 & 1.56 & 1.0 & Dark & 15 \\ 

NGC6822-103 & 296.19068 & $-$14.87268 & 2022 Sep 10 & 960 & 1.39 & 1.1 & Bright & 16 \\ 

NGC6822-175 & 296.22696 & $-$14.80182 & 2022 Oct 11 & 960 & 1.60 & 1.1 & Dark & 13 \\ 
 
WLM~14 & 0.51268 & $-$15.50950 & 2022 Dec 19 & 3360 & 1.40 & 1.2 & Dark & 10 \\
\hline
\end{tabular}
\end{table*}

\section{Molecfit parameters}\label{app:mol_par}
This Appendix provides the optimal \textsc{molecfit} parameters for our EMIR data determined after various tests (Sect.~\ref{sec:tell_corr}).
\begin{table}[h]
    \small
    \centering
    \begin{tabular}{l l }
    \hline
    \hline
      Parameter   & Value \\
      \hline
        LIST\_MOLEC & H20,O2,CO2,CH4,CO \\
   
        FIT\_MOLEC & 1,1,0,0,0 \\
   
        REL\_COL & 1.0,1.0,1.06,1.0,1.0 \\
       
        WLGTOMICRON & 0.0001 \\
        
        VAR\_AIR & VAC \\
        
        FTOL & 1e-10 \\
        
        XTOL & 1e-10 \\
       
        FIT\_RES\_BOX & 0 \\
     
        RELRES\_BOX & 0 \\
       
        FIT\_RES\_GAUSS & 1 \\
       
        RES\_GAUSS & 6.5 \\
       
        FIT\_RES\_LORENTZ & 0 \\
      
        RES\_LORENTZ & 0 \\
        
        FIT\_CONT & 1 \\
        
        CONT\_N & 0 \\
    
        CONT\_CONST & 1 \\
      
        FIT\_WLC & 1 \\
        
        WLC\_N & 0 \\
        
        WLC\_CONST & 0.0038 \\
      
        KERNMODE & 0 \\
  
        KERNFAC & 3 \\
  
        VARKERN & 0 \\
  
        PIXSC & 0.2 \\
        \hline
    \end{tabular}
    \caption{Parameters relevant to the fit carried out by \textsc{molecfit} that are specific to EMIR.}
    \label{tab:mol_par}
\end{table}
\clearpage
\section{Spectral fits}\label{app:fits}

This Appendix provides the results of the fitting process for all the objects except  NGC6822-103 (Figs.~\ref{fit_ngc6822-103} and \ref{plot_ngc6822-103}), which was shown in the main text (Sect.~\ref{specres}).

\begin{figure}[!htb]
   \centering
    \includegraphics[width=0.7\columnwidth]{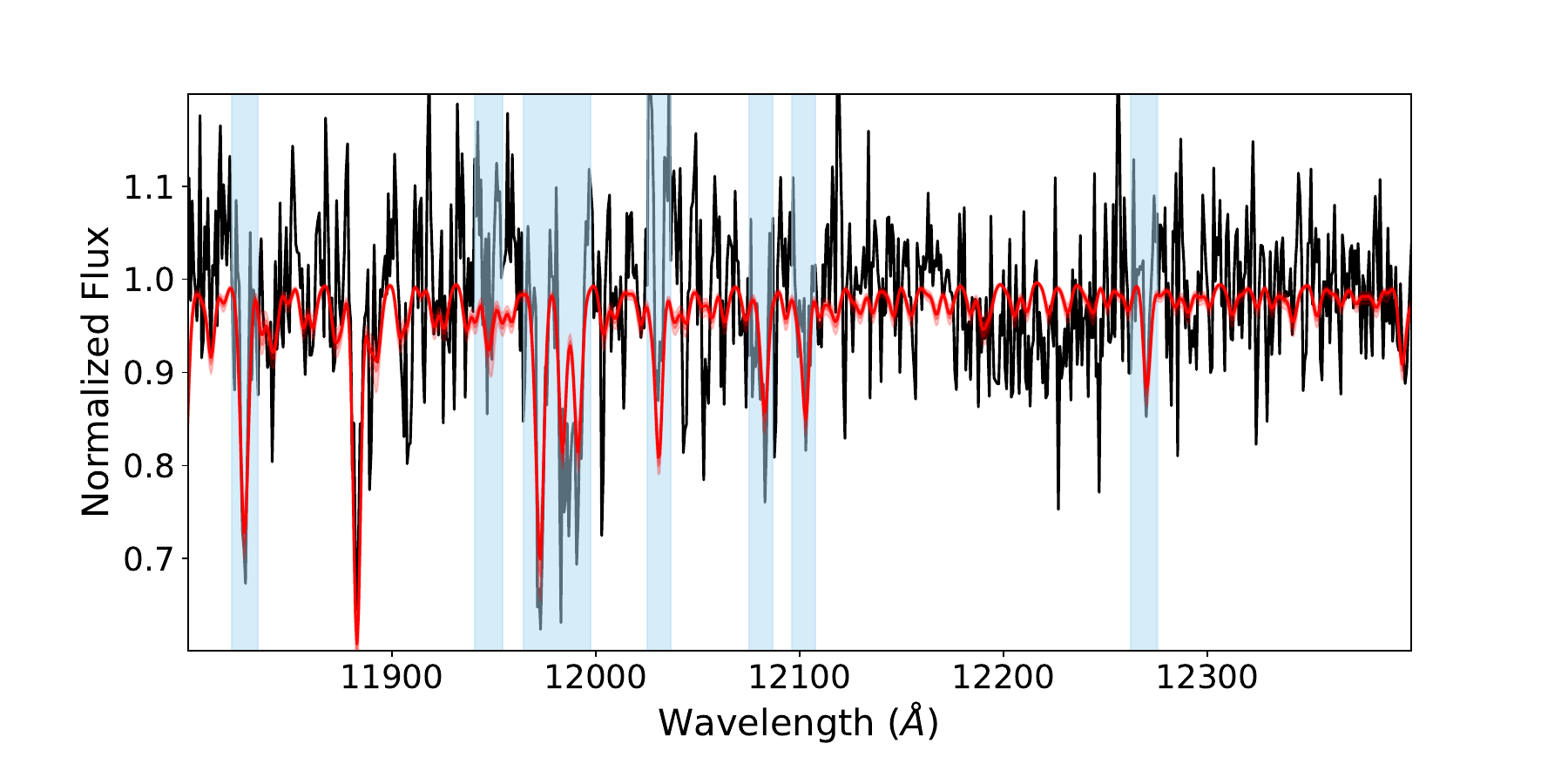}
    \caption{Same as Fig.~\ref{fit_ngc6822-103} for NGC6822-55. Spectral lines Fe~\textsc{I} $\lambda$$\lambda$ 11882.847 and Ti~\textsc{I} $\lambda$$\lambda$ 11892.878 were excluded from the fit, because of artifact contamination.}
    \label{fit_NGC6822-55}
\end{figure}

\begin{figure}[!htb]
   \centering
    \includegraphics[width=0.7\columnwidth]{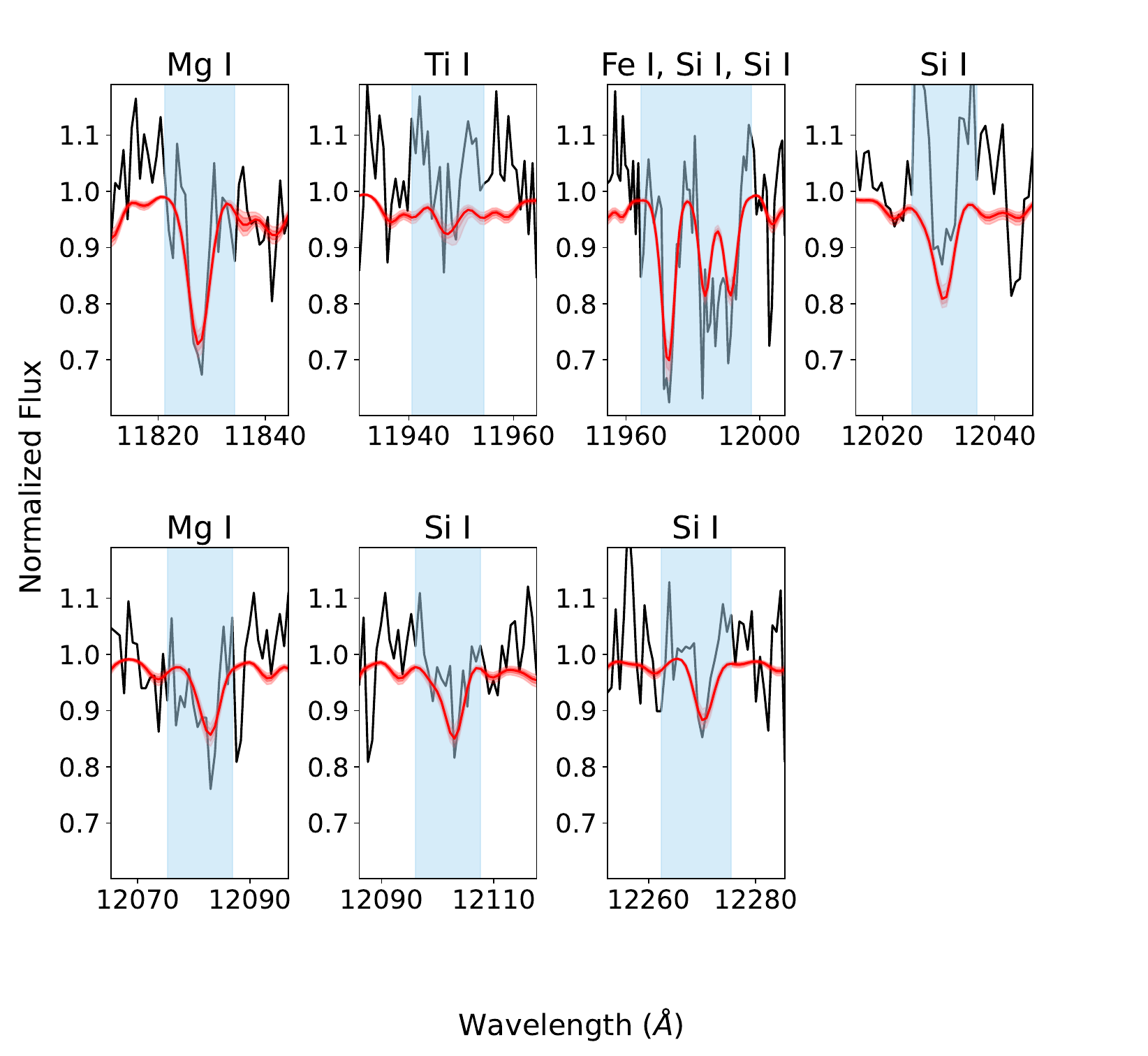}
    \caption{Same as Fig.~\ref{plot_ngc6822-103}, but for NGC6822-55. Spectral lines Fe~\textsc{I} $\lambda$$\lambda$ 11882.847 and Ti~\textsc{I} $\lambda$$\lambda$ 11892.878 were excluded from the fit, because of artifact contamination.}
    \label{plot_NGC6822-55} 
\end{figure}
\clearpage

\begin{figure}
    \centering
    \includegraphics[width=0.7\columnwidth]{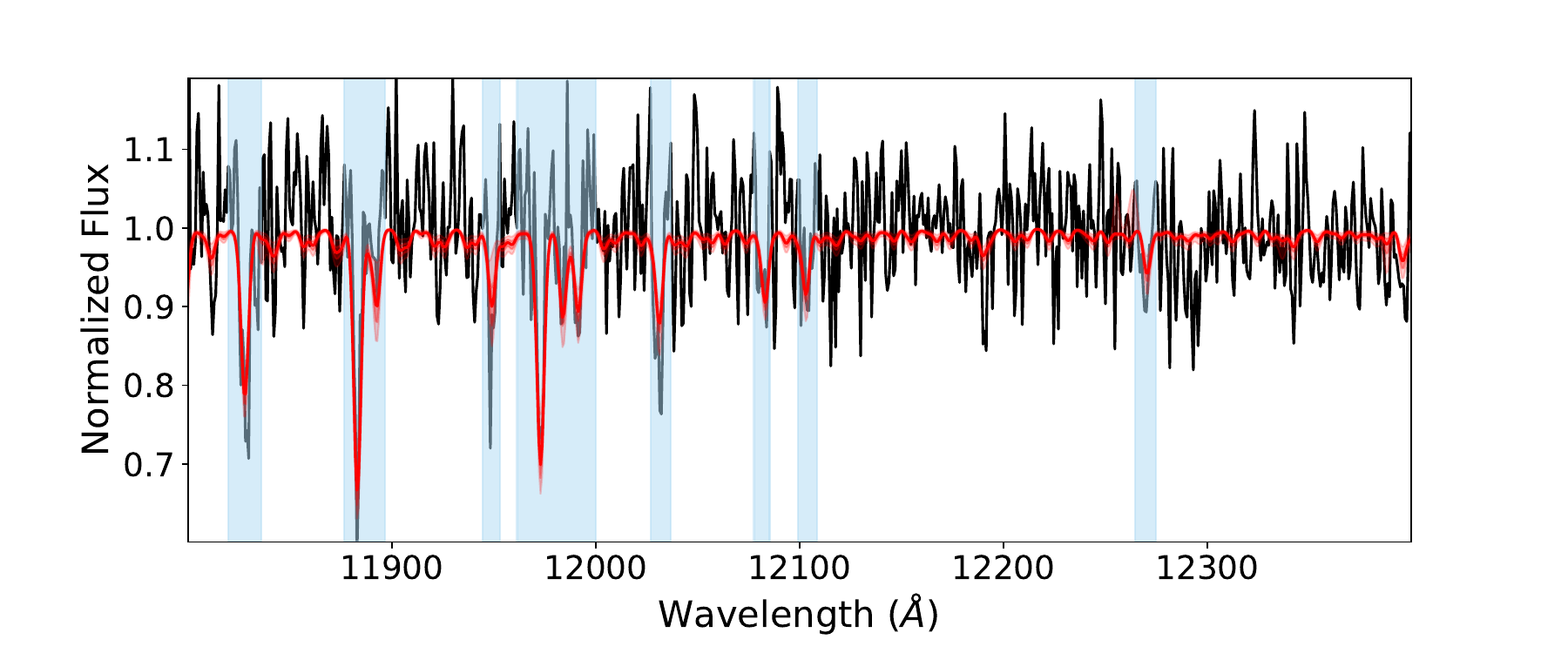}
    \caption{Same as Fig.~\ref{fit_ngc6822-103}, but for NGC6822-70.}
    \label{fit_NGC6822-70}    
\end{figure}

\begin{figure}
    \centering
    \includegraphics[width=0.7\columnwidth]{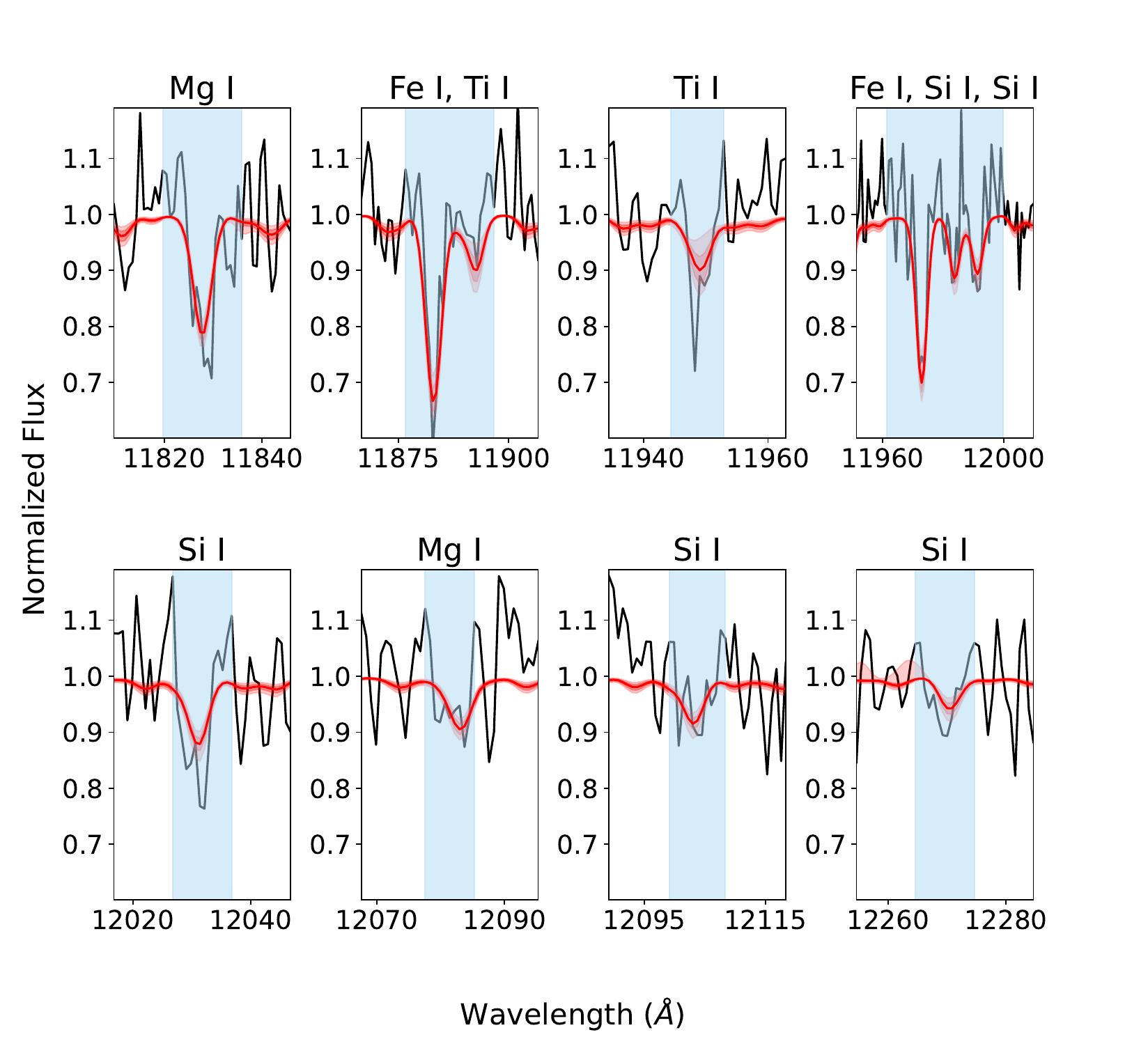}
    \caption{Same as Fig.~\ref{plot_ngc6822-103}, but for NGC6822-70.}
    \label{plot_NGC6822-70} 
\end{figure}

\clearpage
\begin{figure}
    \centering
    \includegraphics[width=0.7\columnwidth]{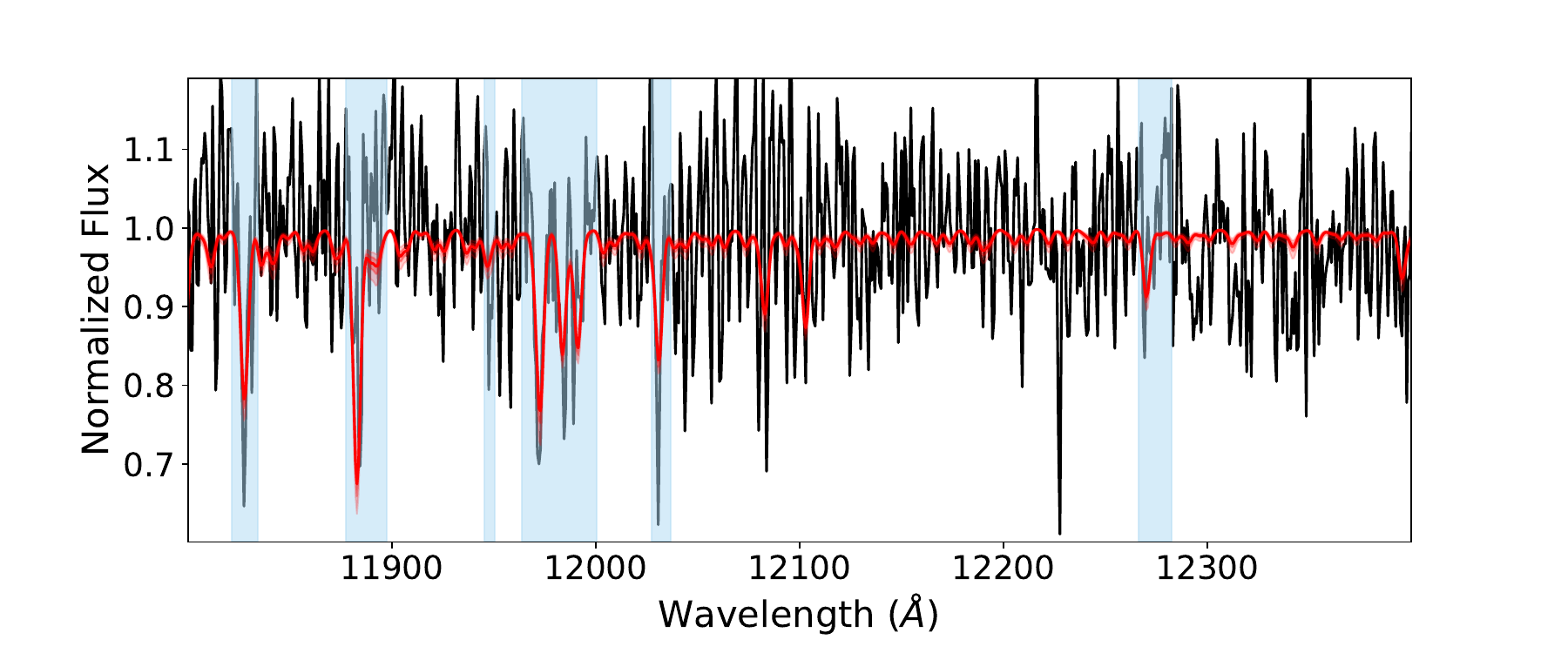}
    \caption{Same as Fig.~\ref{fit_ngc6822-103}, but for NGC6822-175. Spectral lines Si~\textsc{I} $\lambda$ 12103.54 and Mg~\textsc{I} $\lambda$ 12083.346 were excluded from the fit, because their profiles were too noisy.}
    \label{fit_NGC6822-175}    
\end{figure}

\begin{figure}
    \centering
    \includegraphics[width=0.7\columnwidth]{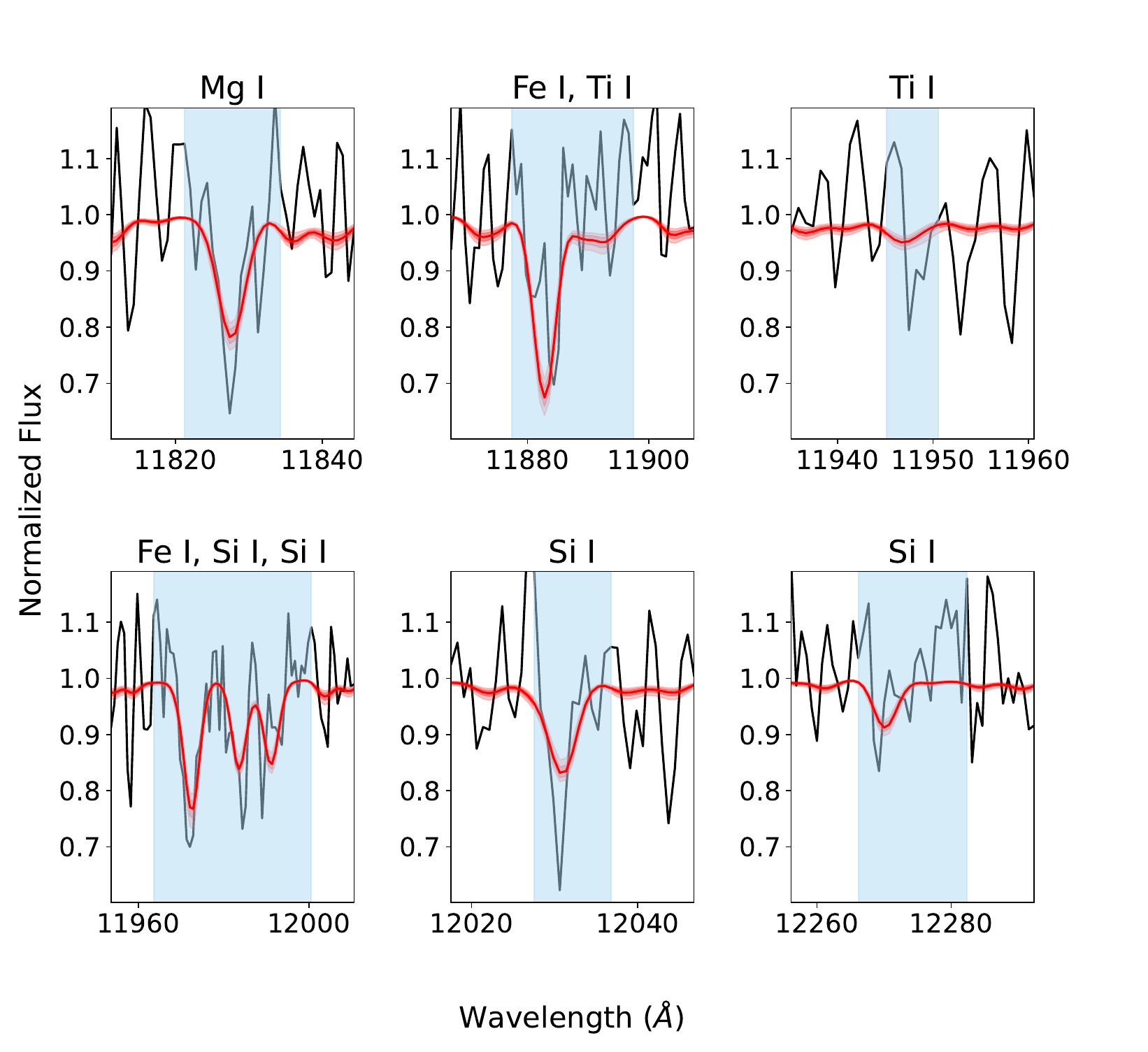}
    \caption{Same as Fig.~\ref{plot_ngc6822-103}, but for NGC6822-175. Spectral lines Si~\textsc{I} $\lambda$ 12103.54 and Mg~\textsc{I} $\lambda$ 12083.346 were excluded from the fit, because their profiles were too noisy.}
    \label{plot_NGC6822-175}
\end{figure}

\clearpage
\begin{figure}
    \centering
    \includegraphics[width=0.7\columnwidth]{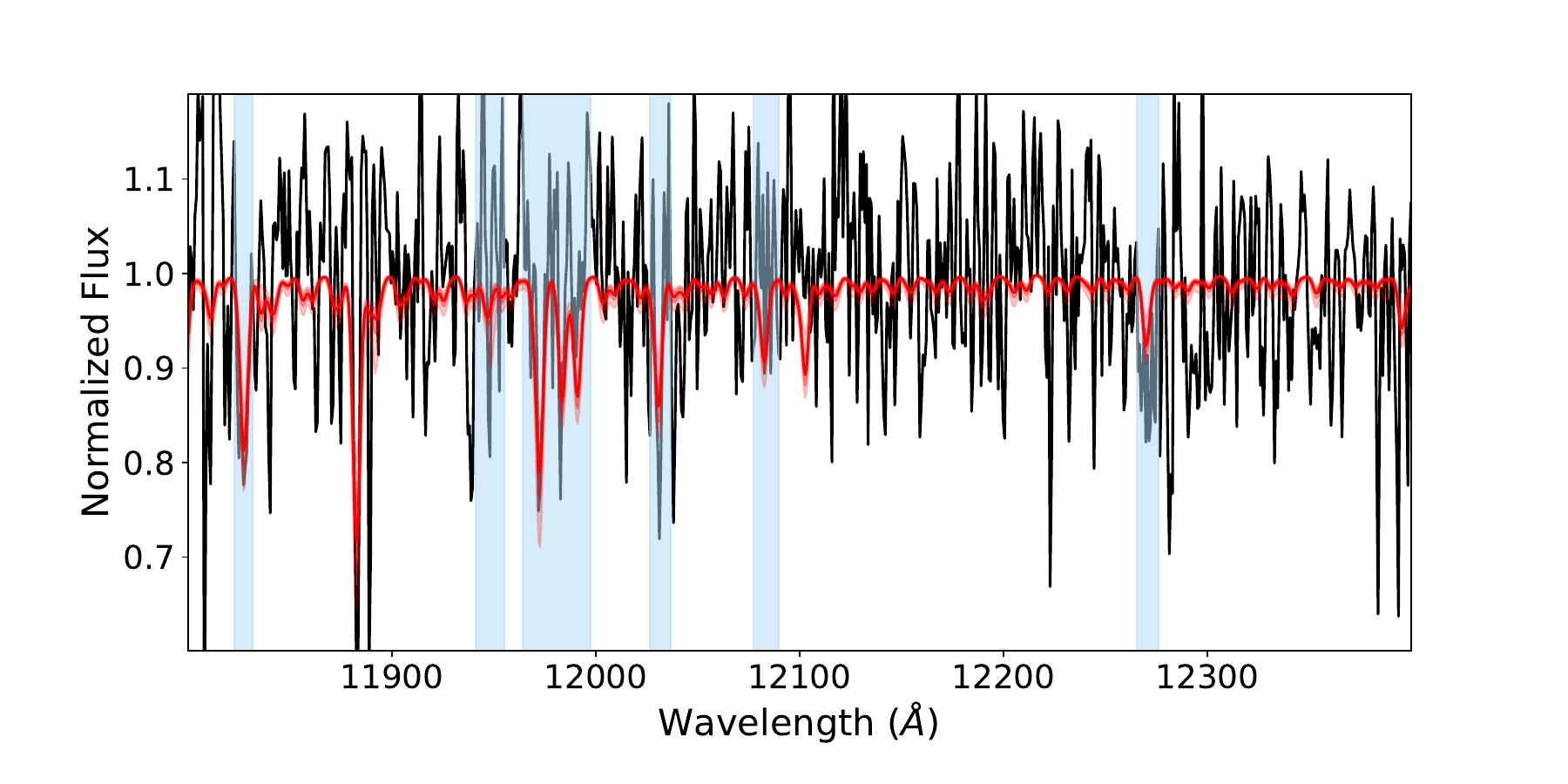}
    \caption{Same as Fig.~\ref{fit_ngc6822-103}, but for WLM~14. Spectral lines Fe~\textsc{I} $\lambda$$\lambda$ 11882.847, Ti~\textsc{I} $\lambda$$\lambda$ 11892.878 were excluded from the fit because of artifact contamination, and Si~\textsc{I} $\lambda$ 12103.54, was excluded because its profile was too noisy.}
    \label{fit_WLM_14}    
\end{figure}

\begin{figure}
    \centering
    \includegraphics[width=0.7\columnwidth]{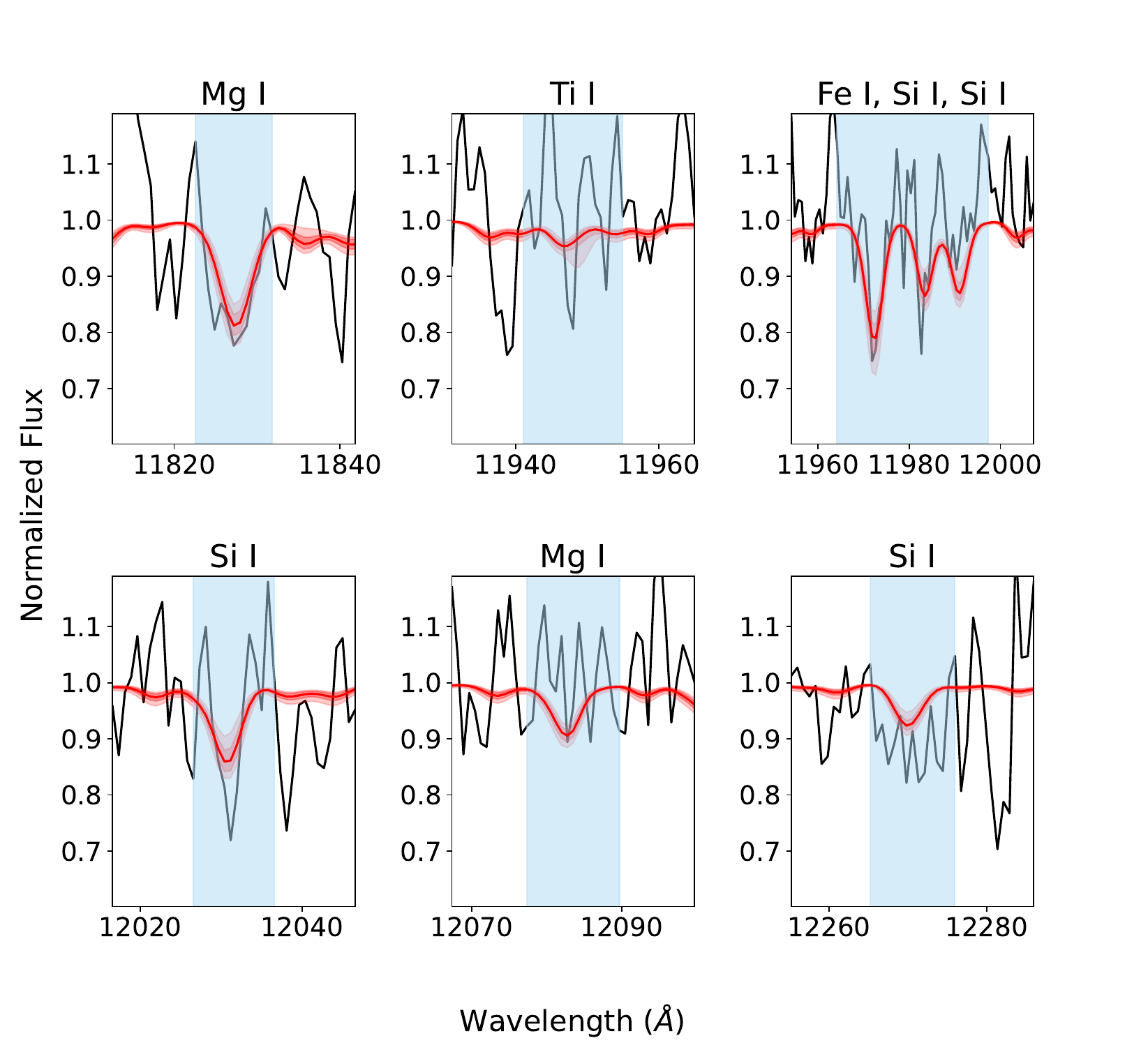}
    \caption{Same as Fig.~\ref{plot_ngc6822-103}, but for WLM~14. Spectral lines Fe~\textsc{I} $\lambda$$\lambda$ 11882.847, Ti~\textsc{I} $\lambda$$\lambda$ 11892.878 were excluded from the fit because of artifact contamination, and Si~\textsc{I} $\lambda$ 12103.54 was excluded because its profile was too noisy.}
    \label{plot_WLM_14} 
\end{figure}

\clearpage
\section{Light curves} \label{app:lc}
This Appendix presents the optical, near-IR, and mid-IR light curves of the targets except for NGC6822-52, which was shown in the main text (Sect.~\ref{sec:lc}). Symbols are described in Fig.~\ref{lc_ngc6822-52}.

\begin{figure*}[h]
\includegraphics[width=1.0\columnwidth]{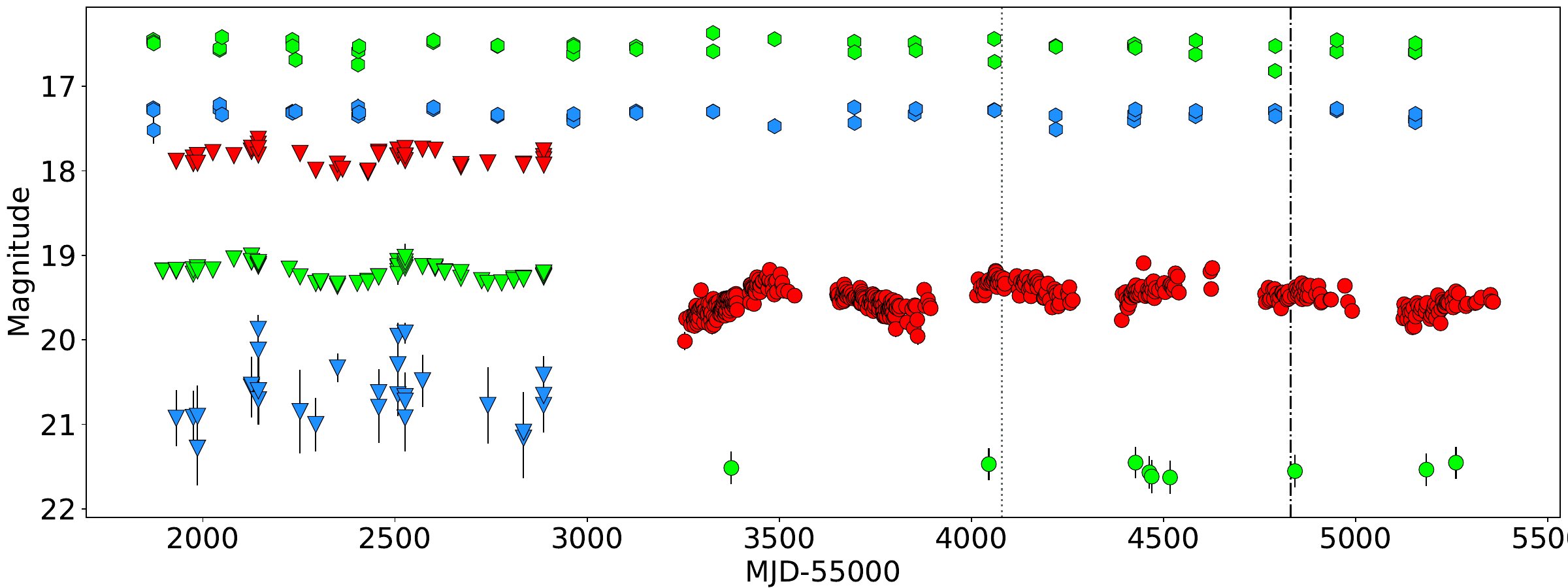}
    \caption{Same as Fig.~\ref{lc_ngc6822-52}, but for IC10-26089 (K4~I). For illustration purposes we show $W1$ $+$ 4.0, $W2$ $+$ 5.0.}
    \label{lc_ic1026089}    
\end{figure*}

\begin{figure*}[h]
    \centering
    \includegraphics[width=1.0\columnwidth]{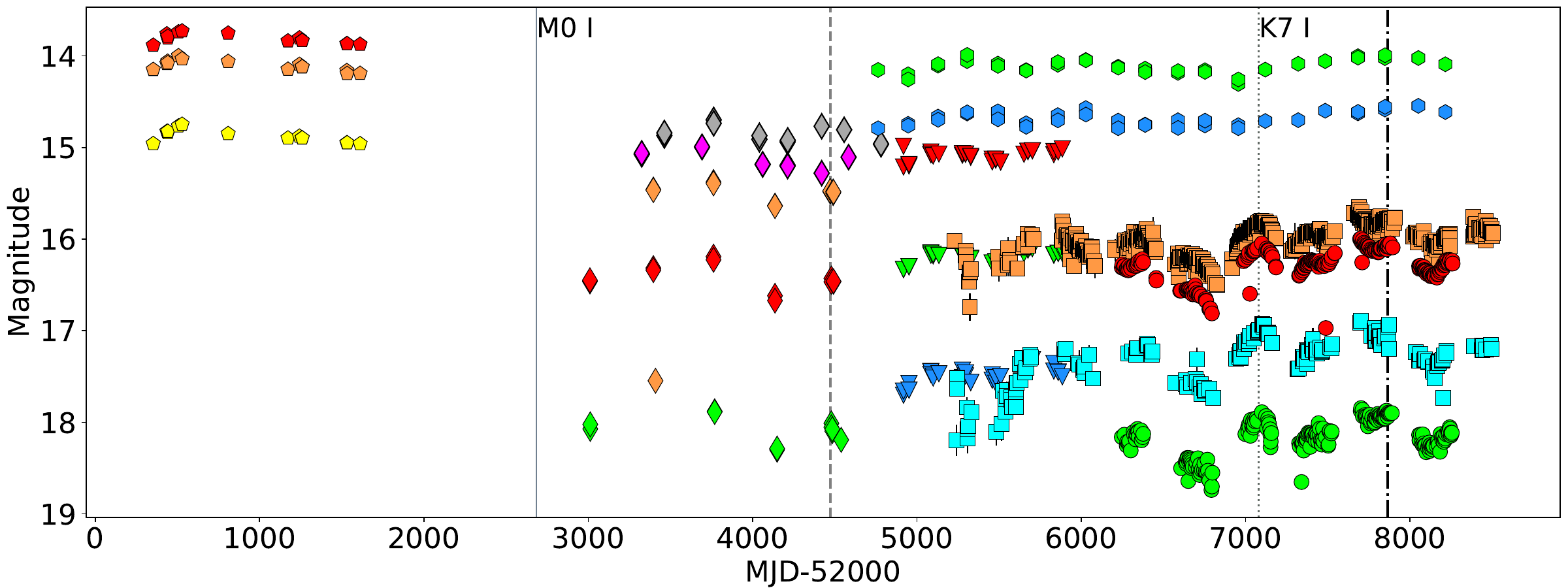}
    \caption{Same as Fig.~\ref{lc_ngc6822-52}, but for NGC6822-55. The gray dashed line denotes the epoch of the spectrum of \citet{Patrick2015}. The spectral type of this object, both at the \citet{Levesque2012} epoch and the OSIRIS epoch, is noted, highlighting its transition to warmer temperatures (see text for more details).}
    \label{lc_ngc6822-55}    
\end{figure*}

\begin{figure*}
    \centering
    \includegraphics[width=1.0\columnwidth]{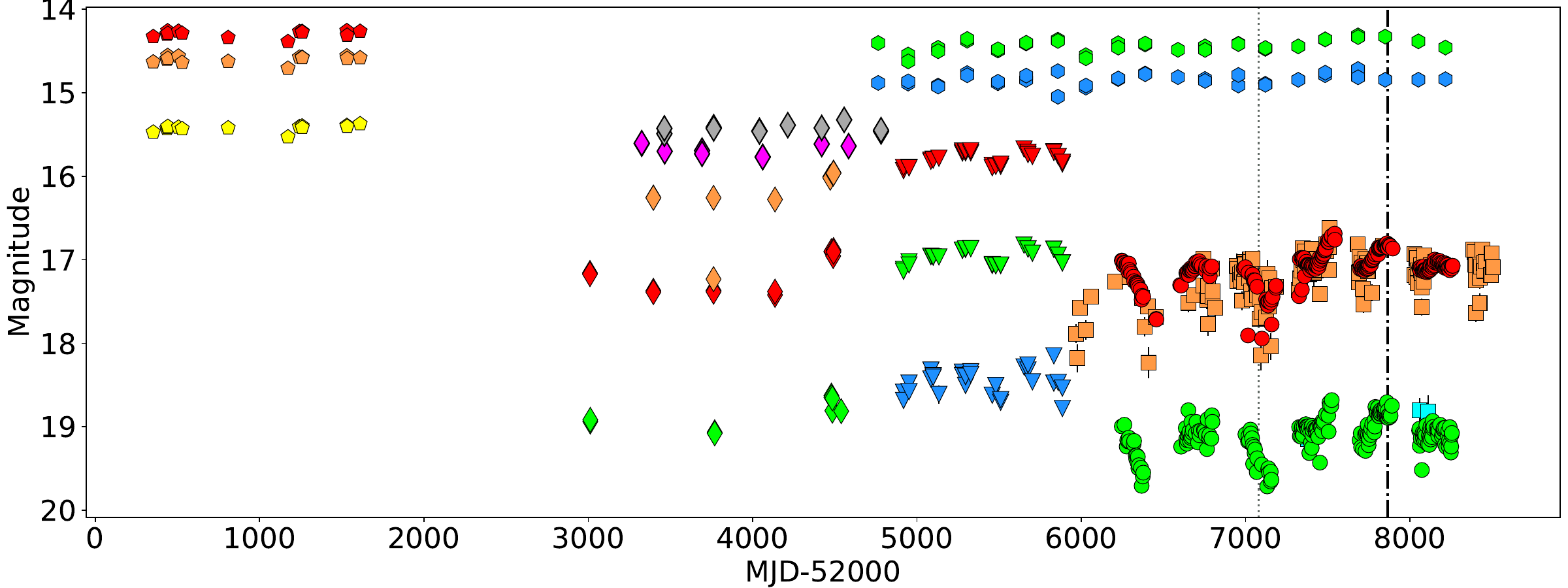}
    \caption{Same as Fig.~\ref{lc_ngc6822-52}, but for NGC6822-70.}
    \label{lc_ngc6822-70}    
\end{figure*}

\begin{figure*}[h]
    \centering
    \includegraphics[width=1.0\columnwidth]{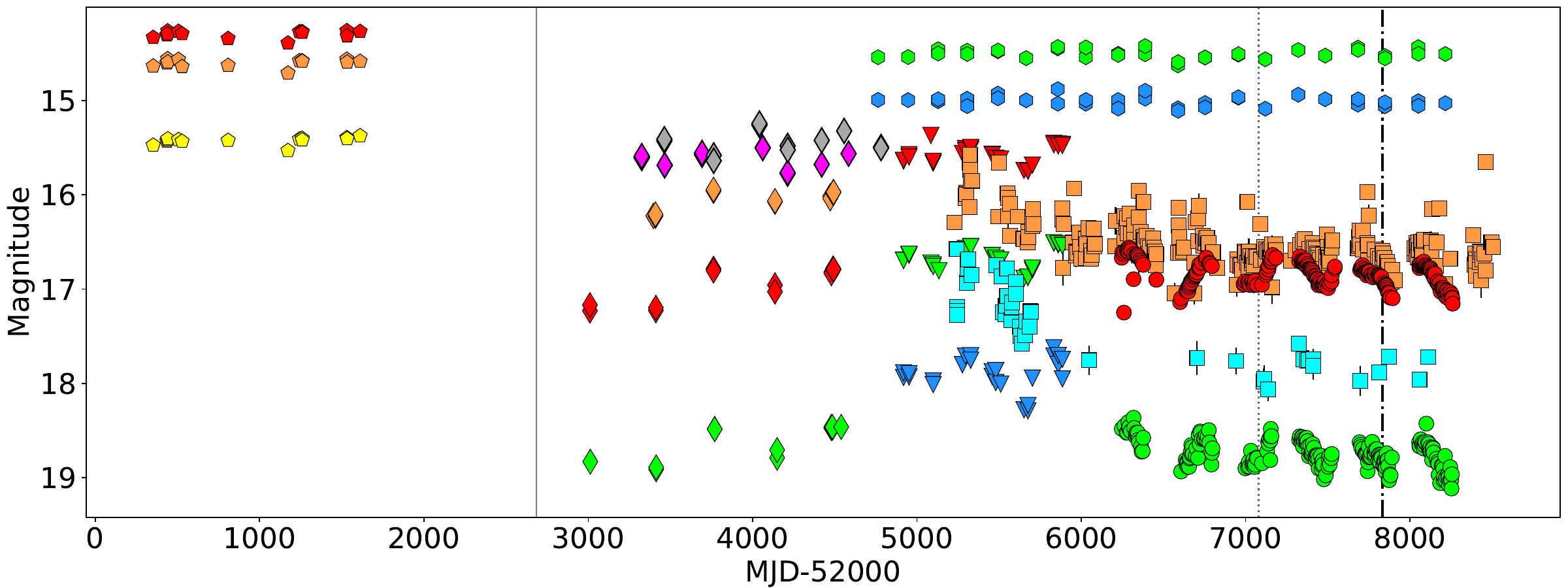}
    \caption{Same as Fig.~\ref{lc_ngc6822-52}, but for NGC6822-103. The gray solid line represents the epoch when the spectrum of \citet{Levesque2012} was obtained.}
    \label{lc_ngc6822-103}    
\end{figure*}

\begin{figure*}[h]
    \centering
    \includegraphics[width=1.0\columnwidth]{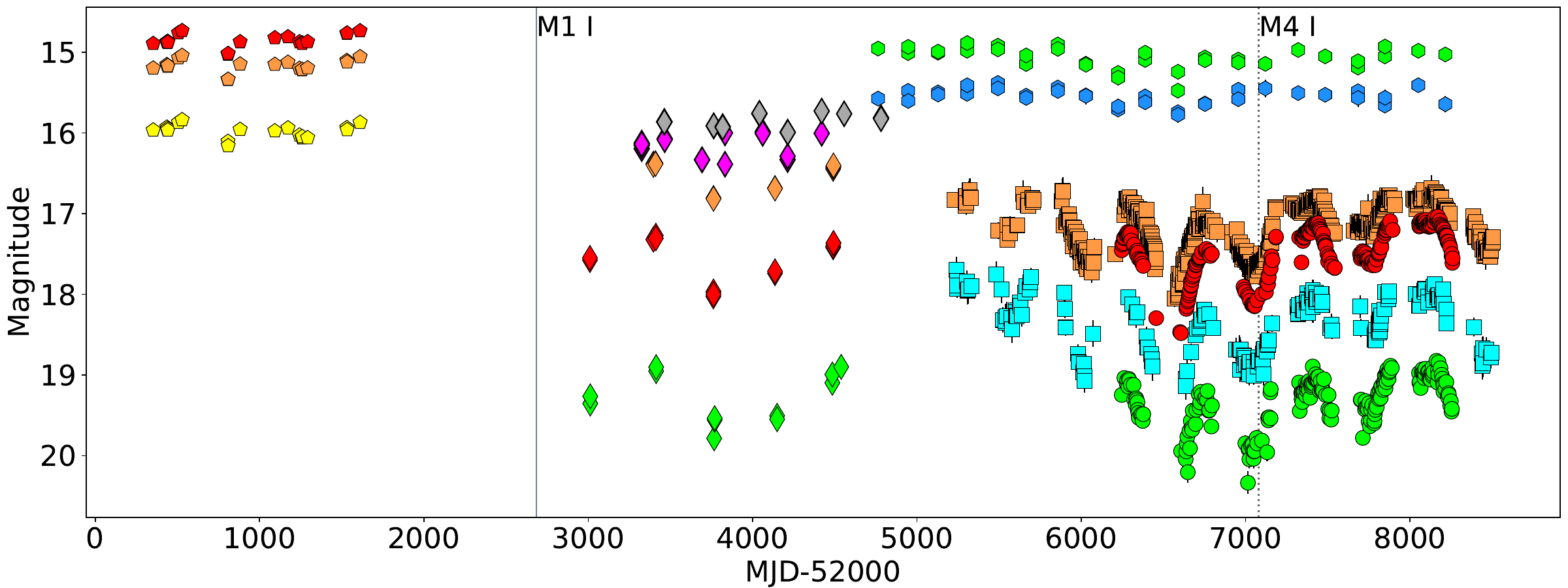}
    \caption{Same as Fig.~\ref{lc_ngc6822-52}, but for NGC6822-175. Lines are as in Fig.~\ref{lc_ngc6822-55}. The spectral type of this object, both at the \citet{Levesque2012} epoch and the OSIRIS epoch, is noted, marking its spectral variability.}
    \label{lc_ngc6822-175}    
\end{figure*}

\begin{figure*}
    \centering
    \includegraphics[width=1.0\columnwidth]{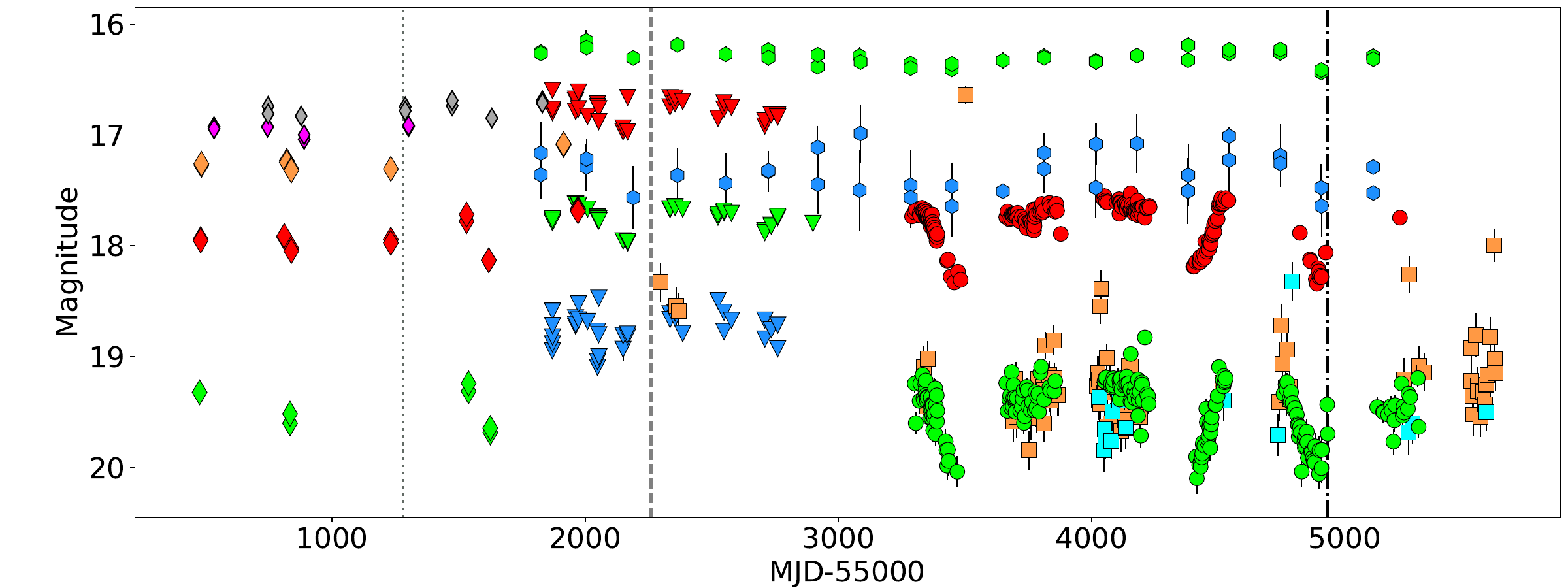}
    \caption{Same as Fig.~\ref{lc_ngc6822-52}, but for WLM~14 (K4-5~I). For illustration purposes we show $W1$ $+$ 2.0, $W2$ $+$ 3.0. The gray dashed line denotes the epoch when the data of \citet{Britavskiy2019} were obtained, while the gray dotted line indicates the epoch that the observations reported in \citet{Britavskiy2015} were carried out.}
    \label{lc_wlm_14}    
\end{figure*}

\clearpage
\section{Periodograms} \label{app:periodograms}
This Appendix provides the periodograms constructed for the 3 sources that display periodic behavior, namely NGC6822-52, WLM~14, and IC10-26089 (Sect.~\ref{sec:phot_res}).

\hfill
\begin{figure*}[h]
   \begin{subfigure}{0.5\columnwidth}
    \centering
    \includegraphics[width=1\linewidth]{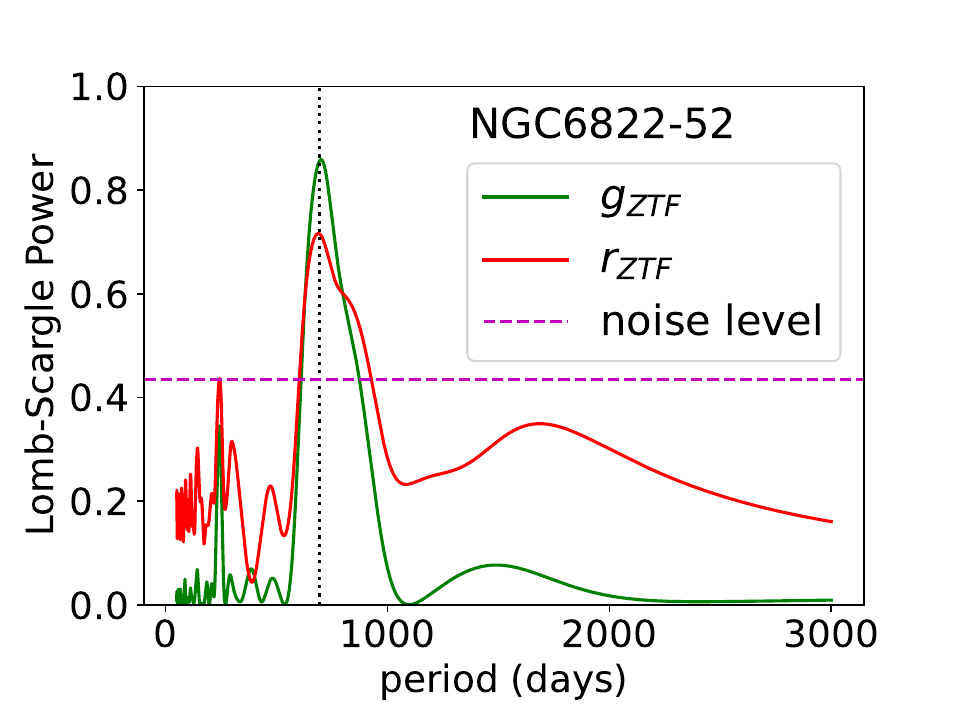}
   \end{subfigure}
\hfill % <--- 
    \begin{subfigure}{0.5\columnwidth}
        \centering
        \includegraphics[width=1\linewidth]{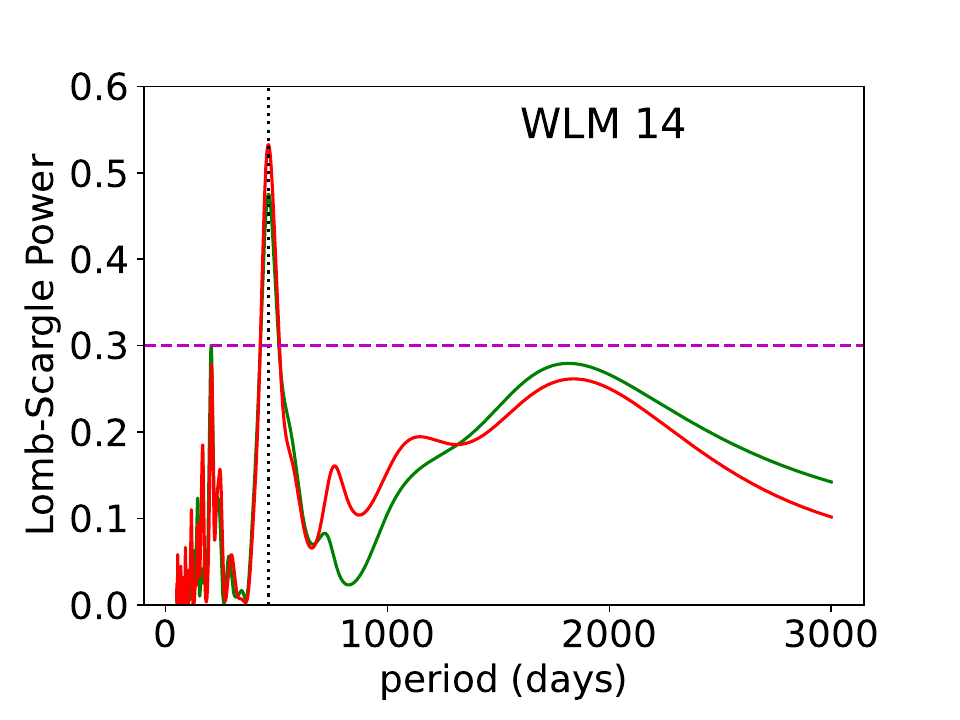}
    \end{subfigure}
\hfill % <--- 
    \begin{subfigure}{0.5\columnwidth}
        \centering
        \includegraphics[width=1\linewidth]{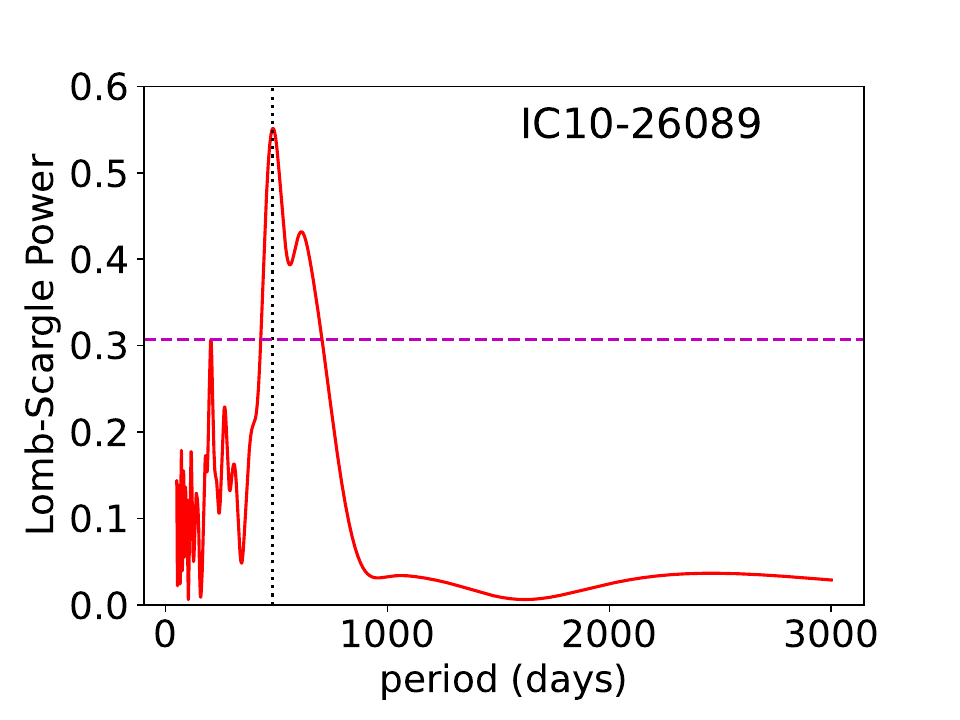} 
    \end{subfigure}
\caption{Periodograms of NGC6822-52 (top left), WLM~14 (top right) and IC10-26089 (bottom). The measured period is indicated by the black-dotted vertical line.}
\label{fig:periodograms} 
\end{figure*}
\end{appendix}
\end{document}